\documentclass[a4paper,11pt]{article}
\pdfoutput=1 

\usepackage{jheppub} 
\usepackage{slashed}
\usepackage{mathtools}
\usepackage{float}
\usepackage[T1]{fontenc} 
\usepackage{color}
\usepackage{amsmath}
\usepackage{amssymb}
\usepackage[extra]{tipa}

\title{\boldmath Inferring the covariant $\Theta$-exact noncommutative coupling in the top quark pair production at linear colliders}

\author[a]{J. Selvaganapathy}
\author[a]{Partha Konar}
\author[b]{and Prasanta Kumar Das}

\affiliation[a]{Theoretical Physics Group, Physical Research Laboratory,\\Ahmedabad,380 009, India}
\affiliation[b]{Department of Physics, BITS-Pilani,Goa campus,\\Goa,403 726, India}

\emailAdd{jselva@prl.res.in}
\emailAdd{konar@prl.res.in}
\emailAdd{pdas@goa.bits-pilani.ac.in}

\abstract{
A novel non-minimal interaction of neutral right-handed fermion and abelian gauge field in the covariant $\Theta$-exact noncommutative 
standard model (NCSM) which is invariant under Very Special Relativity (VSR) Lorentz subgroup, opens an avenue to study the top quark pair 
production at linear colliders. Here the non-minimal coupling is denoted as $\kappa$ and the noncommutative (NC) scale $\Lambda$. 
In this work, we consider two types of analysis, one is without considering helicity basis technique and another, considering helicity 
of the initial and final state particles. Further, the realistic electron and positron beam polarization are taken into account to measure 
the NC parameters. In the first case, when $\kappa$ is positive and certain values of $\Lambda$, we found that a specific threshold value of 
machine energy (optimal energy in the units of GeV) $\sqrt{s_0}$ ($ \simeq 2.52 ~\Lambda + 39$ ) may be quite useful to look for the signature 
of the spacetime noncommutativity with unpolarized beam. The statistical $\chi^2$ analysis of the azimuthal anisotropy which is due to broken 
rotational invariance about the beam axis, is quite possible when $\kappa$ takes negative value $0 >\kappa> -0.596$ which persuade a lower 
limit on NC scale $\Lambda$ ($ 1.0\, \text{to} \, 2.4\,\text{TeV}$) at $\kappa_{max}=-0.296$ with $95\%$ C.L according to luminosity ranging 
from $100\,fb^{-1} \text{to}1000\,fb^{-1}$ at machine energy $\sqrt{s}=1.4\,\text{TeV}\,\text{and}\, 3.0\,\text{TeV}$. In another case, 
we perform detailed analysis for the polarized and unpolarized electron-positron beam to probe spacetime noncommutativity in light of following 
observables like azimuthal anisotropy, helicity correlation, and top quark helicity left-right asymmetry. The polarization of the initial 
beam $\{ P_{e^{-}},P_{e^{+}}\} = \{-0.8,0.3\}( \{-0.8,0.6\})$ enhances the ranges of lower limit 
on $\Lambda$, $i.e.\, 1.13 \, \text{to} \, 2.80\,\text{TeV}$ at $\kappa_{max}$ alongside the $\kappa_{max}$ enhanced 
into $-0.5445 \,(-0.607)$ $95\%$C.L accord with luminosity and machine energy. Finally, we studied the intriguing mixing of the UV and the IR by 
invoking a specific structure of noncommutative anti-symmetric tensor $\Theta_{\mu\nu}$ which is invariant under 
translational $T(2)$ VSR Lorentz subgroup. \\
} 

\keywords{UV/IR mixing, Very special relativity (VSR), Top quark pair production, 
Optimal collision energy, Azimuthal anisotropy and Helicity correlation.}

\begin{document} 
\maketitle

\def\issue(#1,#2,#3){{\bf #1}, #2 (#3)} 

\def\APP(#1,#2,#3){Acta Phys.\ Polon.\ \issue(#1,#2,#3)}
\def\ARNPS(#1,#2,#3){Ann.\ Rev.\ Nucl.\ Part.\ Sci.\ \issue(#1,#2,#3)}
\def\CPC(#1,#2,#3){Comp.\ Phys.\ Comm.\ \issue(#1,#2,#3)}
\def\CIP(#1,#2,#3){Comput.\ Phys.\ \issue(#1,#2,#3)}
\def\EPJC(#1,#2,#3){Eur.\ Phys.\ J.\ C\ \issue(#1,#2,#3)}
\def\EPJD(#1,#2,#3){Eur.\ Phys.\ J. Direct\ C\ \issue(#1,#2,#3)}
\def\IEEETNS(#1,#2,#3){IEEE Trans.\ Nucl.\ Sci.\ \issue(#1,#2,#3)}
\def\IJMP(#1,#2,#3){Int.\ J.\ Mod.\ Phys. \issue(#1,#2,#3)}
\def\JHEP(#1,#2,#3){JHEP\ \issue(#1,#2,#3)}
\def\JPG(#1,#2,#3){J.\ Phys.\ G \issue(#1,#2,#3)}
\def\MPL(#1,#2,#3){Mod.\ Phys.\ Lett.\ \issue(#1,#2,#3)}
\def\NP(#1,#2,#3){Nucl.\ Phys.\ \issue(#1,#2,#3)}
\def\NIM(#1,#2,#3){Nucl.\ Instrum.\ Meth.\ \issue(#1,#2,#3)}
\def\PL(#1,#2,#3){Phys.\ Lett.\ \issue(#1,#2,#3)}
\def\PRD(#1,#2,#3){Phys.\ Rev.\ D \issue(#1,#2,#3)}
\def\PRL(#1,#2,#3){Phys.\ Rev.\ Lett.\ \issue(#1,#2,#3)}
\def\PTP(#1,#2,#3){Progs.\ Theo.\ Phys. \ \issue(#1,#2,#3)}
\def\RMP(#1,#2,#3){Rev.\ Mod.\ Phys.\ \issue(#1,#2,#3)}
\def\SJNP(#1,#2,#3){Sov.\ J. Nucl.\ Phys.\ \issue(#1,#2,#3)}
\def\PR(#1,#2,#3){Phys.\ Rep.\ \issue(#1,#2,#3)}

\flushbottom

\section{Introduction}
\label{sec:intro}
  After the discovery of the Higgs boson (the only missing link of the standard model) at the Large Hadron Collider (LHC) at CERN,
  the Standard Model (SM) of Particle Physics, despite its inability of explaining the 
  higgs hierarchy problem, neutrino mass problem, baryon asymmetry etc, is now widely believed 
  to be a low energy effective field theory. Now if there is physics beyond the standard model (BSM), 
  it is expected to show up at the TeV energy colliders i.e. at the LHC or at the upcoming electron-positron Linear Collider. Among the class of 
  BSM models, supersymmetry, extra dimension, space-time noncommutativity are found to 
  be quite interesting because of their rich phenomenological content. A host of phenomenological investigations 
  have already been made and are available in the literature.  
  \par  The gravity which becomes strong at the TeV energy scale in large extra dimensional 
  model \cite{ADD,AADD}, makes the spacetime noncommutative (i.e. fuzzy) at the TeV scale.
  The signature of the TeV scale spacetime noncommutativity can be found at the TeV 
  energy electron-positron linear collider or hadron colliders. Snyder's pioneering work 
  \cite{Snyder} and the recent advancement in low energy string theory manifests the fact that 
  spacetime can be noncommutative \cite{CDS,DH,SW,SST} and it has generated a lot of 
  enthusiasm about the TeV scale spacetime noncommutativity among the particle physics community.   
   In 1996, Witten et.al., \cite{Witten,HW} has suggested that one can probe the stringy effects by 
   lowering the threshold value of noncommutativity to TeV, a scale which is not so far from 
   present or future collider energy scale. 
  \par Now at high energy when gravity becomes strong, the spacetime coordinates 
   become an operator $\hat{x}_\mu$. They no longer commute, satisfying the algebra 
   \begin{equation}
   [\hat{x}_\mu,\hat{x}_\nu]=i\Theta_{\mu\nu} =  \frac{i c_{\mu\nu}}{\Lambda^2} \label{XXTheta} 
   \end{equation}
   Here $\Theta_{\mu\nu}$ (of mass dimension $-2$) is real and antisymmetric tensor.
   $c_{\mu\nu}$ is the antisymmetric constant and $\Lambda$, the noncommutative scale. 
   In the noncommutative space, the ordinary product between fields is replaced by 
   Moyal-Weyl(MW) \cite{DN,RJ,JS} star($\star$) product defined by
   \begin{equation}
   (f\star g)(x)=exp\left(\frac{i}{2}\Theta_{\mu\nu}\partial_{x^\mu}\partial_{y^\nu}\right)f(x)g(y)|_{y=x}.
   \label{StarP}
   \end{equation}
   In the theoretical point of view, it is expected that the spacetime noncommutative (NC) field theories should 
   reduce into a commutative field theory when $\Theta\rightarrow0$ or whenever the momenta of the field quanta are much smaller than $1/\sqrt{|\Theta|}$.
   Although such naive expectation is valid at tree level interaction but it suffers by infamous UV/IR mixing phenomenon at the loop level which is 
   shown by Filk and Minwalla in \cite{filk,YMNC1,Minwalla,IRseiberg1} respectively. The UV/IR mixing arises due to non-planar Feynman diagram at loop level, 
   precisely due to Moyal phase. But in the NC gauge theory, the expansion of Moyal star product between fields breaks the invariance of the truncated action.
   Seiberg and Witten \cite{SW} formulated a map as a power series of $\Theta$ which relates the NC gauge theory and commutative gauge theory, thereby
   the NC action is gauge invariant at any order $\Theta$. There are two types of map which are, the expansion of $\Theta$ with keeping gauge field all order and
   the expansion of gauge field with keeping $\Theta$ all order namely $\Theta$-expanded and $\Theta$-exact Seiberg-Witten map (SW map) respectively.
   The $\Theta$-expanded SW map for matter field $\psi$, gauge field $A^\mu$ are written as a power series of the noncommutative parameter $\Theta$ as follows \cite{DH,SW,Witten,HW,DN,RJ,JS}
   \begin{eqnarray}
   \widehat{\psi}(x,\Theta) & = & \psi(x) + \Theta \psi^{(1)} + \Theta^2 \psi^{(2)} + \cdots \\
   \widehat{A}_\mu(x,\Theta) & = & A_\mu(x) + \Theta A_\mu^{(1)} + \Theta^2 A_\mu^{(2)} + \cdots \label{SWM} 
   \end{eqnarray}
   The second order ($e^{2}$), third order($e^{3}$) and further expansions of $\Theta$-exact SW maps are rather complicated than $\Theta$-expanded SW map 
   construction due to gauge structure degrees of freedom \cite{Trampetic3}. In the $\Theta$-expanded approach, the NC theory is renormalizable in linear order 
   at one-loop level \cite{Minwalla}. In the $\Theta$-exact SW map approach, the UV/IR mixing is related to phase factor of distinctive star product between commutative fields.
   The phenomenological implications of the $\Theta$-exact SW map and properties of UV/IR mixing pursued extensively in ref \cite{Trampetic4,Martin1}.
   There are enormous amount of work has been done considering Moyal-Weyl (MW) approach \cite{Minwalla,MWURIR1,MWURIR2,MWURIR3,MWURIR4,MWURIR5,MWURIR6} as well as Seiberg-Witten Map approach
   \cite{SWURIR1,SWURIR2,SWURIR3,SWURIR4,SWURIR5,SWURIR6,SWURIR7,SWURIR8,SWURIR9,HKT3} in order to remove the UV/IR mixing. The role of supersymmetric Yang-Mills theory 
   in the noncommutative UV/IR mixing \cite{petrov1,petrov2,Trampetic1,Trampetic2,ABJM} were extensively scrutinized to overcome the divergences which appeared in the field non-local interaction.
   Other class of NC theories namely, non-geometric theories like $\kappa$-Minkowski spacetime and Snyder spacetime also exhibits the UV/IR mixing at one-loop level \cite{kappa1,Snyder1,Snyder2}.
   In general, the UV/IR mixing is the universal property of all NCQFT. But in the ref \cite{kappa2}, it is shown that the truncated $\kappa$-deformed action does not possess the UV/IR mixing due to implementation of the 
   $\kappa$-deformed star products in the $\kappa$-Minkowski spacetime which evades the integral measure problems. On the contrary, in the 
   Snyder-type star product in the Snyder spacetime, the $\phi^{4}$ interaction breaks the translational invariance as well as it celebrates
   the UV/IR mixing at the one-loop tadpole contribution to the two-point function \cite{Snyder1,Snyder2}. The non-associativity induces such UV/IR mixing
in the Synder NC scalar field theory in both momentum-conserving as well as momentum-nonconserving approach.
   Moreover notwithstanding the UV/IR mixing, the covariant $\Theta$-exact NC theory \cite{HKT1} has cosmological implications on the decoupling/coupling temperature of right-handed neutrino
   at the early universe \cite{ptolemy1,ptolemy2,astro2020} as well as primordial nucleosynthesis \cite{HT} and ultra high energy (UHE) cosmic ray experiments \cite{HKT2}.
   \par In the phenomenological point of view, based on the MW approach, the Moller scattering, Bhabha scattering and electron-photon scattering were first studied in ref \cite{HPR, Hewett}. 
   The processes $\gamma \gamma \to e^+ e^-$ and  $\gamma \gamma \to \gamma \gamma$ were investigated 
   in \cite{Mathews,Rizzo}. A review on NC phenomenology is available here \cite{HKM}. 
   A lot of NC phenomenology using the Seiberg-Witten technique is available. 
   Calmet et.al., \cite{CJSWW,CW} first constructed the standard model in the noncommutative 
   spacetime, which is the minimal version of the Noncommutative Standard Model (mNCSM). 
   There exists another version, called non-minimal version of NCSM i.e. nmNCSM \cite{Melic} in which 
   besides the usual mNCSM interaction vertices, a host of new vertices e.g. triple neutral gauge boson vertices also arises which are found to be absent in the 
   mNCSM and of course in the SM. These new vertices lead to several interesting decays 
   e.g. $Z \to \gamma \gamma,~ g g$ which are forbidden in the SM. These were studied in \cite{BDDSTW,AJSW,DST,BLRT1,EH1}. 
   In the SW map approach, ref \cite{AOR} first investigated the neutral vector boson ($\gamma$, $z$) pair production 
   in the noncommutative spacetime at the LHC and they obtained 
   the bound on the noncommutative scale $\Lambda \ge 1~\rm{TeV}$. 
   The $W^\pm$ pair production \cite{Ohl} also studied and found that the  azimuthal distribution is oscillatory 
   corresponding to $\Lambda_{NC} = 0.7~\rm{TeV}$, which differs significantly from the flat distribution obtained in the SM.
   Recently, in \cite{AEN} the single top quark production at LHC
   found that the cross section deviates significantly from the standard model 
   for $\Lambda_{NC} \ge 0.98~\rm{TeV}$. We first studied \cite{SELVA1} the impact of mNCSM 
   on the top quark pair production in the expanded SW map approach at electron-positron linear collider considering the effect of earth rotation. 
   The impact of mNCSM on Higgstralung process in linear collider studied by us using $\Theta$-exact SW map in \cite{SELVA2}. 
   For the first time, the Drell-Yan process in the nmNCSM, we investigated the exotic photon-gluon-gluon and 
   $Z$-boson-gluon-gluon interaction at the LHC and obtained $\Lambda_{NC} \ge 0.4~\rm{TeV}$ \cite{SELVA3}. 
  In the present work we study the effect of spacetime noncommutativity on the pair production 
  of top quarks production using helicity technique in the TeV energy electron-positron collider within the context 
  of covariant $\Theta$-exact NCSM. We predominantly focusing on certain center of mass energy of the electron-positron collision which is future plan of the CLIC\cite{CLIC1,CLIC2}.
  \par The content of the paper is as follows. In Sec.\ref{sec:covariant}, we briefly describe the covariant 
   $\Theta$-exact NCSM and the Light-like noncommutativity from 
   Cohen-Glashow's Very Special Relativity (VSR) theory. We further obtain the cross-section 
   of the top quark pair production in $e^-~ e^+$ collision in the $\Theta$-exact NCSM. 
   In Sec.\ref{sec:inferk}, we studied the cross-section of the top quark pair production in $e^{+}\,e^{-}$ collision by constructing the NC observables
   $\Delta \sigma$ and obtaing bounds on them, finding thus the optimal collision energy, for a given $\kappa$ and $\Lambda$ respecticely,
   at which one should look for the signature for spacetime noncommutativity.
   We also discuss the azimuthal anisotropy and investigate its sensitivity on the NC scale $\Lambda$ and the non-minimal coupling $\kappa$. 
   We performed detailed helicity analysis for $t \overline{t}$ production at Sec.\ref{sec:calcul2}. From top quark and anti-top quark helicity correlation and left-tight top asymmetry, we obtain constrain on 
   $\Lambda$ for different $\kappa$ values, we also performed the polarized beam 
   analysis to obtain lower bound on $\Lambda$. In Sec.\ref{sec:conclude}, we summarize and conclude.
   In appendix, we have shown that the removal of UV/IR mixing and UV divergence free neutrino self energy correction in the $\Theta$-exact NCSM by choosing VSR T(2) invariant $\Theta_{\mu \nu}$.  
  Finally, we have presented the matrix element expressions for top pair production with and without considering the helicity amplitude analysis.
 \section{Covariant $\Theta$-exact Noncommutative Standard Model (NCSM)}\label{sec:covariant}
The construction of NCSM is based on enveloping the $SU(N)$ Lie 
algebra via Seiberg-Witten map. The SW map is a map between the NC fields and commutative spacetime fields which satisfies the gauge equivalence principle and gauge consistency principle.
The power series expansion of the noncommutative tensor ($\Theta^{\mu \nu}$) known as $\Theta$-expanded SW map and power series expansion of the gauge field 
($V^{\mu}$) namely $\Theta$-exact SW map. In the $\Theta$-expanded SW map approach, the fields are expanded like order by order $\Theta$ but it perpetuate 
in all order gauge field. Similarly, in the $\Theta$-exact SW map, the fields are
expanded order by order gauge field but it accommodate all order $\Theta$ terms.
The $\Theta$-exact SW map has few advantages while deriving tree-level SM field interaction by 
keeping terms up to the order $\mathcal{O}(V^{\mu})$ as well as the renormalized
field, coupling and mass up to one loop.
In the NCSM, the Higgs scalar field $\widehat{\Phi}$ is a functional of two gauge fields and it transforms covariantly under gauge transformation.
Eventually, Higgs scalar field takes hybrid SW map and hybrid gauge transformation to preserve the gauge covariant Yukawa terms in the Yukawa sector.
So the Higgs scalar field can couple with left-handed and right-handed SM fermions via the "left" charge and the "right" 
charge of the fermion. However, the scalar field is not a different particle, in contrast, it has different NC representation of 
SW map\cite{MelicEW}. Extending this approach to the gauge sector as well as fermion sector, we get the hybrid SW map for fermion fields and gauge fields\cite{HKT1,STWR,HT,HKT2}.
One can derive the $\Theta$-exact hybrid SW map for NC fields which are given as follows,\\
\textit{Fermion (lepton) hybrid Seiberg-Witten map:}
\begin{eqnarray}
 \widehat{\Psi}_{L} & = & \Psi_{L} - \frac{\Theta^{\mu \nu}}{2}\left( g\, A^{a}_{\mu} T^{a} + Y_{L}\, g_{Y} B_{\mu} \right)\bullet \partial_{\nu} \Psi_{L} 
 - \Theta^{\mu \nu} \, \kappa \, g_{Y} \, B_{\mu} \circledast \partial_{\nu} \Psi_{L} + \mathcal{O}(V^{2}) \Psi_{L} \nonumber \\
 \widehat{l}_{R} & = & l_{R} - \frac{\Theta^{\mu \nu}}{2}(Y_{R}\, g_{Y} B_{\mu} )\bullet \partial_{\nu} l_{R} 
 - \Theta^{\mu \nu} \, \kappa \, g_{Y} \, B_{\mu} \circledast \partial_{\nu} l_{R} + \mathcal{O}(V^{2}) l_{R} \nonumber \\
 \widehat{\nu}_{R} & = & \nu_{R} - \Theta^{\mu \nu} \, \kappa \, g_{Y} \, B_{\mu} \circledast \partial_{\nu} \nu_{R} + \mathcal{O}(V^{2}) \nu_{R} \label{hybridswm}
\end{eqnarray}
The gauge transformations are defined as 
\begin{eqnarray}
 \delta_{\widehat{\Lambda}}\left( \begin{array}{c}
\widehat{\nu}_{L}  \\
\widehat{l}_{L}
\end{array} \right) & = & i \,g_{Y} \,\left[( Y_{L}+\kappa )\widehat{\Lambda}\star \left( \begin{array}{c}
\widehat{\nu}_{L}  \\
\widehat{l}_{L}
\end{array} \right) - \kappa \left( \begin{array}{c}
\widehat{\nu}_{L}  \\
\widehat{l}_{L}
\end{array} \right) \star \widehat{\Lambda} \right] \nonumber \\
 \delta_{\widehat{\Lambda}} \widehat{l}_{R} & = & i \,g_{Y} \,\left[(Y_{R}+\kappa)\widehat{\Lambda}\star \widehat{l}_{R} - \kappa \, \widehat{l}_{R} \star \widehat{\Lambda} \right] \label{gaugetrans}  
\end{eqnarray}
Here $\kappa$ is the fermion non-minimal coupling, $\widehat{\Lambda}$ is NC gauge parameter and $Y_{L},Y_{R}$ are lepton hypercharges and $g,g_{Y}$ are the $SU(2)_{L},U(1)_{Y}$ gauge couplings,
respectively. The gauge potential $V^{\mu}$, defined by the SM group $G_{SM}= U(1)_{Y}\otimes SU(2)_{L}\otimes SU(3)_{C}$, can be written as 
$$ V_{\mu}(x)= g_{Y}\,Y B_{\mu}(x)+ g\,\sum_{a=1}^{3}T^{a}_{L}A_{\mu}^{a}(x)+ g_{s}\sum_{b=1}^{8}T^{b}_{s}G_{\mu}^{b}(x).$$
The products (used above) are defined as
\begin{equation}
 f\, \star\, g = f \left( e^{\frac{i}{2}\overleftarrow{\partial}_{\mu} \Theta^{\mu \nu} \overrightarrow{\partial}_{\nu}}\right)  g \, ;
 \quad  f\, \bullet \, g = f \left( \frac{e^{\frac{i}{2}\overleftarrow{\partial}_{\mu} \Theta^{\mu \nu} \overrightarrow{\partial}_{\nu}} -1} {\frac{i}{2} \overleftarrow{\partial}_{\mu} \Theta^{\mu \nu} \overrightarrow{\partial}_{\nu}} \right)  g \, ;
 \quad  f\, \circledast \, g = f \left( \frac{\sin\left(\frac{1}{2}\overleftarrow{\partial}_{\mu} \Theta^{\mu \nu} \overrightarrow{\partial}_{\nu} \right)} {\frac{i}{2} \overleftarrow{\partial}_{\mu} \Theta^{\mu \nu} \overrightarrow{\partial}_{\nu}} \right)  g \label{starprod} 
 \end{equation}
In the equation(\ref{starprod}), $\Theta^{\mu \nu}$ is the constant antisymmetric tensor
which can have arbitrary structure. Upon imposing some special spacetime group symmetry 
such, one can get particular conserved symmetry quantity which may reduce certain degrees
of freedom in the choice of representation of $\Theta^{\mu \nu}$.

\subsection{Light-like noncommutativity from Cohen-Glashow VSR theory }\label{sec:vsr}
The Poincare group symmetry is the fundamental symmetry in high energy particle physics which is postulated by special theory of 
relativity. At high energies (say over the Planck energy), it is believed that the usual description of the space-time is typically no 
longer valid, the Lorentz symmetry is violated over the minimal length scale. 
Hence one cannot have full Lorentz symmetry group at such minimal length scale, which 
is required to be extended at such minimal length scale (high energy).\\
In the minimal Cohen-Glashow Very Special Relativity (VSR) theory\cite{CGVSR}, 
the VSR subgroups are defined under certain symmetry with space-time translation and 
2-parametric proper subgroup of Lorentz group $SO(3,1)$ 
whose generators are $K_{x}+J_{y}$ and 
$K_{y}-J_{x}$. Here $\bf{K}$ and $\bf{J}$ are the generators of boost and rotation, respectively.\\
The representation of the VSR subgroups are the representation of the Lorentz group
instinctively  but not vice-versa. The VSR subgroups are $T(2), \, E(2), \, HOM(2)$ and $SIM(2)$, 
respectively. Here, $T(2)$ is the $2$-parametric translation group on two dimensional plane. The 
group generators are $T_{1}=K_{x}+J_{y}$ and $T_{2}=K_{y}-J_{x} $ which satisfy $[T_{1}\, , \,T_{2}]=0$.
The subgroup $ E(2)$ is the $3$-parametric group on two dimensional Euclidean motion with 
$T_{1},\, T_{2}$ and $J_{z}$ as the group generators.  
The subgroup $HOM(2)$ is the group of orientation-preserving similarity invariant group or 
Homotheties group. The generators of $HOM(2)$ group are $T_{1},\, T_{2}$ and $K_{z}$.
Finally, $SIM(2)$ is the $4$-parametric isomorphic group of similitude group and the 
group generators are  $T_{1},\, T_{2}, \, J_{z}$ and $K_{z}$.

 Among the four VSR subgroups, $T(2)$ is the only sub group which admits Lorentz symmetry 
 on the noncommutative tensor $\Theta^{\mu \nu}$ in the Moyal space.
The invariant condition can be written as \cite{CXVSR1,CXVSR2,NCVSR}
\begin{equation}
 \Lambda_{i \, \,\alpha}^{\mu} \Lambda_{i \, \, \beta}^{\nu} \Theta^{\alpha \beta} = \Theta^{\mu \nu}
\end{equation}
where the $T(2)$ subgroup elements are $ \Lambda_{1}=e^{i\xi T_{1}}$ and $ \Lambda_{2}=e^{i \chi T_{2}}$
and the infinitesimal transformation gives
\begin{equation}
 T^{\mu}_{i\, \, \alpha} \Theta^{\alpha \nu}+T^{\nu}_{i\, \, \beta} \Theta^{\mu \beta} = 0 \label{infy}
\end{equation}
One can construct two quantities in the Moyal space which are related to noncommutative 
scale and smallest volume of the spacetime as follows
$$ \zeta^{4} = \Theta_{\mu \nu} \Theta^{\mu \nu} \quad \quad L^{4} = \epsilon^{\alpha \beta \mu \nu} \Theta_{\mu \nu} \Theta_{\alpha \beta} $$
Depending on the choice of $\zeta$ and $L$, we can classify the noncommutativity\cite{Carroll1}.
We are interested in which $ \zeta^{4}=0$ and $L^{4}=0$ namely,
light-like noncommutativity. 
\begin{equation}
  \zeta^{4} = \Theta_{\mu \nu} \Theta^{\mu \nu}\,\equiv\,0;\quad \quad L^{4} = \epsilon^{\alpha \beta \mu \nu} \Theta_{\mu \nu} \Theta_{\alpha \beta} \,\equiv\,0; \label{lightNC}
\end{equation}
Interestingly, the choice of $\zeta^{4}=0$, $\Theta^{\mu \nu}$ 
removes the UV/IR mixing as well as UV divergences typically in the $\Theta$-exact 
SW map noncommutative field theory which we showed in the appendix \ref{app:uvir}.
By imposing the condition(\ref{infy}) and $\zeta^{4}=0$, we get the solution
\begin{equation}
  \Theta^{0i}=-\Theta^{3i} \quad \quad i=1,2.
\end{equation} 
The elements of the $T_{1}$ and $T_{2}$ are $ (T_{1})^{0}_{\,1}=(T_{1})^{1}_{\,0}=(T_{1})^{1}_{\,3}=i,\quad (T_{1})^{3}_{\,1}=-i$ and 
 $ (T_{2})^{0}_{\,2}=(T_{2})^{2}_{\,0}=(T_{2})^{2}_{\,3}=i,\quad (T_{2})^{3}_{\,2}=-i$.
Finally, we obtain the light-like noncommutative antisymmetric tensor ($\Theta_{\mu \nu} \Theta^{\mu \nu} = 0$)as follows
\begin{equation}
 \Theta^{\mu \nu} = \frac{1}{\Lambda^{2}}
\left(\begin{array}{cccc}
0 & -a & -b & 0   \\
a & 0 & 0 & -a \\
b & 0 & 0 & -b \\
0 & a & b & 0 
\end{array} \right) 
\end{equation}
The above $\Theta^{\mu \nu}$ although breaks the rotational invariance but it preserves 
translation symmetry which is invariant under $T(2)$ subgroup.
There are two real free parameters $a$ and $b$ and one can assume $b=0$ (or $a=0$) for the sake of simplicity. 
We have taken the following structure of $\Theta^{\mu \nu}$, which admits azimuthal 
anisotropy by virtue of broken rotational symmetry. Eventually, it reduces the 
computational complexity in the removal of UV/IR mixing in the NC field theory as well which shown in the appendix \ref{app:uvir}.
\begin{equation}
 \Theta^{\mu \nu} = \frac{1}{\Lambda^{2}}\left( \begin{array}{cccc}
0 & -1 & 0 & 0   \\
1 & 0 & 0 & -1 \\
0 & 0 & 0 & 0 \\
0 & 1 & 0 & 0 
\end{array} \right) \label{nctensor}
\end{equation}
The above mentioned antisymmetric tensor (\ref{nctensor}) is the class of light-like, but we 
follows $(0,1,2,3)$ basis instead of light-cone basis$(+,-,1,2)$.

\subsection{Top quark pair production in the Covariant $\Theta$-exact NCSM }\label{sec:calcul1}
The non-zero commutator of the $U(1)_{Y}$ abelian gauge field and SM fermion provides the non-minimal interaction in the background
antisymmetric tensor field. Such type of background tensor interaction equally acts on the 
SM fermion field e.g. quarks, leptons or left handed, right handed particle, irrespective of 
their group representation. In this scenario, although the neutral current interactions get 
modified, the charged current interaction doesn't. The electroweak covariant 
derivative of the $\Theta$-exact NCSM can be written as follows
\begin{eqnarray}
 D_{\mu \, L} \, \widehat{\Psi}_{L} & = & \partial_{\mu \, L}\, \widehat{\Psi}_{L}-i g \, \widehat{A}_{\mu \,L}^{a} T^{a} \star \widehat{\Psi}_{L} - i (Y_{\Psi_{L}}+\kappa)\, g_{Y} \, \widehat{B}_{\mu}\star \widehat{\Psi}_{L}+i \kappa\, g_{Y} \, \widehat{\Psi}_{L} \star \widehat{B}_{\mu}  \label{leftfermion} \\
 D_{\mu \, R} \, \widehat{\Psi}_{R} & = & \partial_{\mu \, R}\, \widehat{\Psi}_{R}- i (Y_{\Psi_{R}}+\kappa)\, g_{Y} \, \widehat{B}_{\mu}\star \widehat{\Psi}_{R}+i \kappa\, g_{Y} \, \widehat{\Psi}_{R} \star \widehat{B}_{\mu} \label{rightfermion}
\end{eqnarray}
where $\kappa$ is the non-minimal coupling of the fermion. 
The equations (\ref{leftfermion},\ref{rightfermion}) contain terms that comprises the neutral fermion 
interaction (gauge-invariant) with the SM photon. A detailed study on 
the photon-neutrino interaction and its related phenomenology can be found in \cite{STWR,HKT3}. 

\noindent One can derive the Feynman rule from the lepton action\cite{HKT1,Wang,wang1,wang2} 
by using equation (\ref{leftfermion},\ref{rightfermion}) and equation (\ref{hybridswm}) as follows
\\
\textit{Photon-Fermion-Fermion interaction:}
\begin{eqnarray}
 ie\, Q_{f}\,\gamma_{\mu}+\frac{e}{2}\left\{i Q_{f} \widetilde{F}_{\bullet}(p_{i},p_{o}) - 2 \, \kappa \, \widetilde{F}_{\circledast}(p_{i},p_{o}) \right\} 
 \left[ (p_{o}\Theta)_{\mu} (\slashed{p}_{i}-m)+ (\slashed{p}_{o}-m)(\Theta p_{i})_{\mu} +(p_{i}\Theta p_{o}) \gamma_{\mu}\right] \nonumber \\ \label{photonfeyn}
\end{eqnarray}
\textit{$Z$ boson-Fermion-Fermion interaction:}
\begin{eqnarray}
 & & \frac{ie}{\sin 2\theta_{w}} \, \gamma_{\mu}(C_{vf}-C_{af}\gamma_{5}) + \nonumber \\  
 &  &  \frac{ie}{\sin 2\theta_{w}} \, \left\{ \frac{i}{2} \widetilde{F}_{\bullet}(p_{i},p_{o}) \right\} \left[ (p_{o}\Theta)_{\mu} (\slashed{p}_{i}-m)+ (\slashed{p}_{o}-m)(\Theta p_{i})_{\mu} +(p_{i}\Theta p_{o}) \gamma_{\mu}\right] (C_{vf}-C_{af}\gamma_{5}) \nonumber \\
 & + & \kappa \, e \, \tan\theta_{w}  \widetilde{F}_{\circledast}(p_{i},p_{o}) \left[ (p_{o}\Theta)_{\mu} (\slashed{p}_{i}-m)+ (\slashed{p}_{o}-m)(\Theta p_{i})_{\mu} +(p_{i}\Theta p_{o}) \gamma_{\mu}\right]  \label{zbosonfeyn}
\end{eqnarray}
Here $ \widetilde{F}_{\bullet}(p_{i},p_{o}) = 2 (\frac{e^{i(p_{i}\Theta p_{o})/2}-1}{i p_{i}\Theta p_{o}}) $ and $ \widetilde{F}_{\circledast}(p_{i},p_{o}) = 2 (\frac{\sin((p_{i}\Theta p_{o})/2)}{p_{i}\Theta p_{o}}) $, 
$p_{i}$ and $p_{o}$ are the ingoing and outgoing fermion momenta towards vertex. 
Also $ p_{i}\Theta p_{o} = p_{i}^{\mu} \Theta_{\mu \nu} p_{o}^{\nu}$ and $\theta_{w}$ is the Weinberg angle. \\
Let us consider the top-quark pair production process in electron-positron which proceeds
via the $S$-channel exchange of $\gamma$ and $Z$ bosons i.e. 
$ e^{-}e^{+} \xrightarrow{\gamma/Z} t \overline{t}$.
\begin{figure}[h]
\centering 
\includegraphics[width=1.0\textwidth]{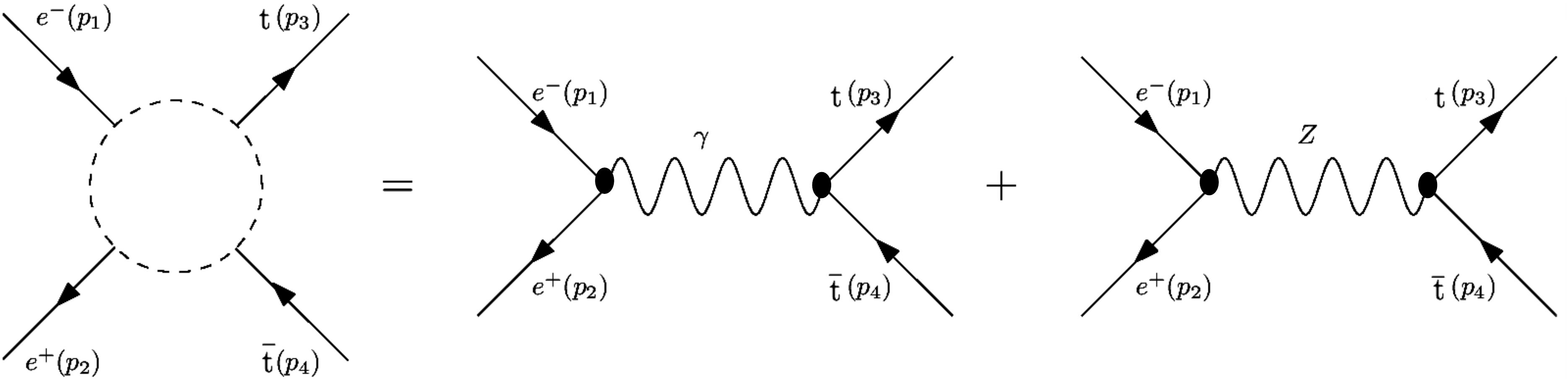}
\caption{\label{fig:1} Feynman diagram for top quark pair production at linear colliders}
\end{figure}
The processes we consider are the tree-level processes and the external particles are on-shell. 
Applying the equation of motion to the external particles and using $\Theta_{\mu\nu}$ (given in equation \ref{nctensor}), 
the Feynman rules are simplified as  
\begin{eqnarray}
 i \,e\, Q_{e} \gamma_{\mu} \, \, \label{photon1}\\ 
\frac{i\,e}{\sin2\theta_{w}}\gamma_{\mu} (C_{ve}-C_{ae}\gamma_{5})  \label{zboson1}
\end{eqnarray}
for $\gamma-e-e$ and $Z-e-e$ interaction vertices, respectively. Because of the specific choice 
of $\Theta_{\mu\nu}$ (equation \ref{nctensor}), we find $p_{1}\Theta p_{2}=0$ and other 
momentum dependent simply vanishes by virtue of the equation of motion and the result turns 
out to be equivalent to the SM interaction. Here $Q_{e}$ is the electron charge and 
$C_{ve}$ and $C_{ae}$ are the vector and the axial-vector coupling of the electron with the $Z$ boson. 
On the other hand $ p_{4}\Theta p_{3} \neq 0$, and we get 
\begin{eqnarray}
 i \,e\, Q_{t} \gamma_{\mu} \,e^{i\frac{p_{4}\Theta p_{3}}{2}}- 2 \, \kappa \,e \, \gamma_{\mu}\sin\left(\frac{p_{4}\Theta p_{3}}{2}\right) \label{photon2}\\
 \frac{i\,e}{\sin2\theta_{w}}\gamma_{\mu} (C_{vt}-C_{at}\gamma_{5})\,e^{i\frac{p_{4}\Theta p_{3}}{2}}+ 2 \, \kappa \,e \, \tan\theta_{w} \, \gamma_{\mu}\sin\left(\frac{p_{4}\Theta p_{3}}{2}\right) \label{zboson2}
\end{eqnarray}
for $\gamma-t-t$ and $Z-t-t$ vertices respectively. Here $Q_{t}$ is the top quark charge 
and $C_{vt}$ and $C_{at}$ are the vector and the axial-vector couplings of the top quark 
with the $Z$ boson. Note that $C_{vf} = T_{3f} - 2 Q_f sin^2 \theta_w$ and $C_{af} = T_{3f}$, where $f = e, t$. \\ 

\noindent{\textbf{Top pair differential cross section }}\\
The top pair production proceeds via the $s$ channel exchange of photon($\gamma$) and $Z$ boson 
and the tree level Feynman diagrams are shown in the Fig.\ref{fig:1}. 
The total noncommutative differential cross section  can be written as
$$  \frac{d\sigma^{NC}}{d\Omega} = \frac{d\sigma^{\gamma }}{d\Omega} + \frac{d\sigma^{Z}}{d\Omega} + \frac{d\sigma^{\gamma Z}}{d\Omega}. $$
The detailed calculations of NC differential cross section are given in the appendix \ref{app:ncxsec}. 
The signature of the noncommutativity arises due to the term  
 $\frac{p_{4}\Theta p_{3}}{2} = \left( \frac{s\, \beta}{4 \Lambda^{2}} \right) \sin\theta \, \cos\phi$. 
 Here $\theta$ is the polar angle which is defined with respect to the electron beam axis i.e along the $z$-axis
 and $\phi$ is the azimuthal angle. Since $\frac{d\sigma^{NC}}{d\Omega}$ is proportional to the momentum $\Theta$ weighted product
$p_{4} \Theta p_{3}$, the azimuthal distribution possibly throw some light for probing the fermion non-minimal coupling 
$\kappa$ as a noncommutative signal.
\section{Inferring the $ \Theta$-exact NC coupling $\kappa$ and NC scale $\Lambda$}\label{sec:inferk}
\begin{figure}[h]
\centering 
\includegraphics[width=.5\textwidth]{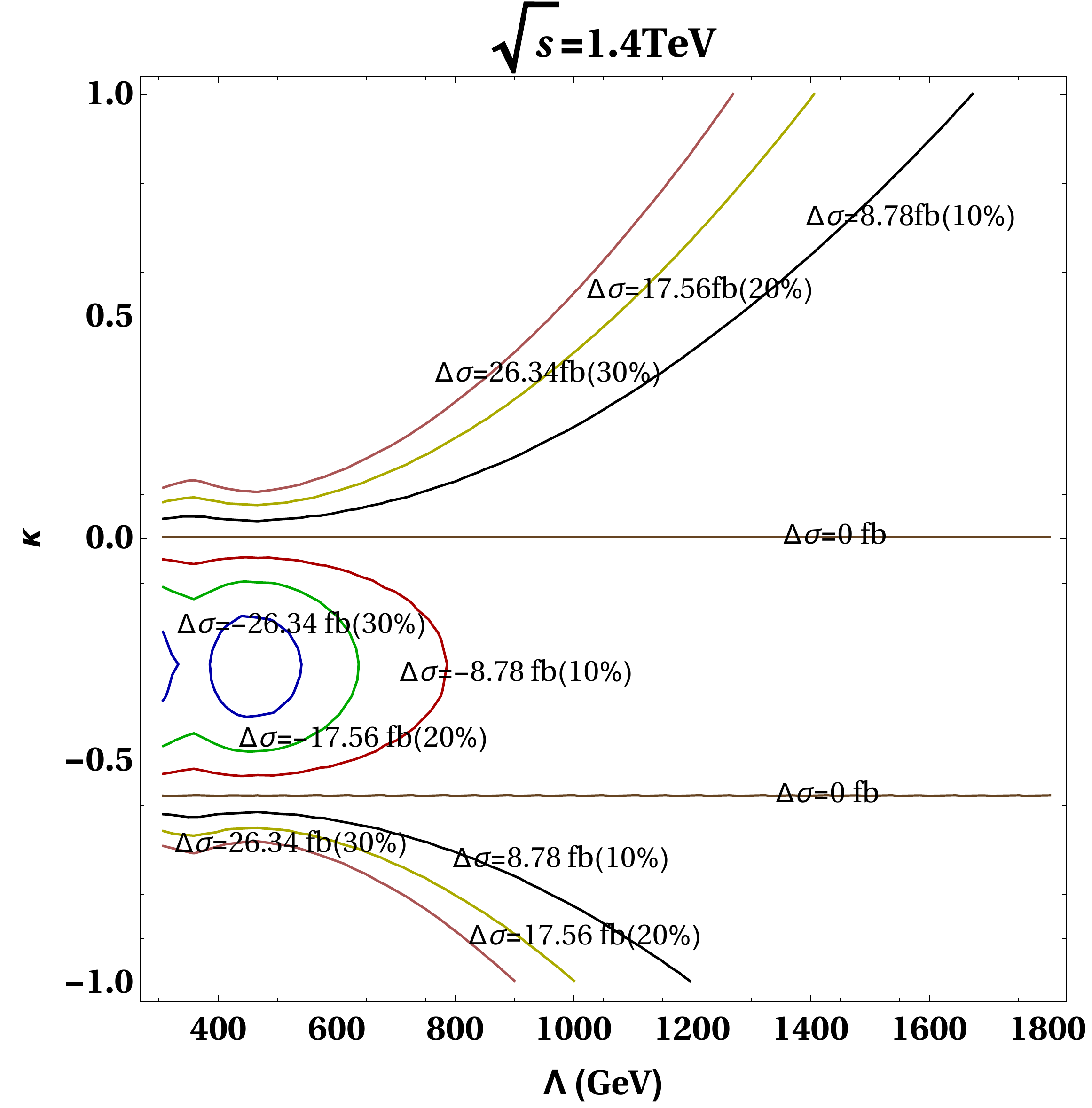}
\hfill
\includegraphics[width=.5\textwidth]{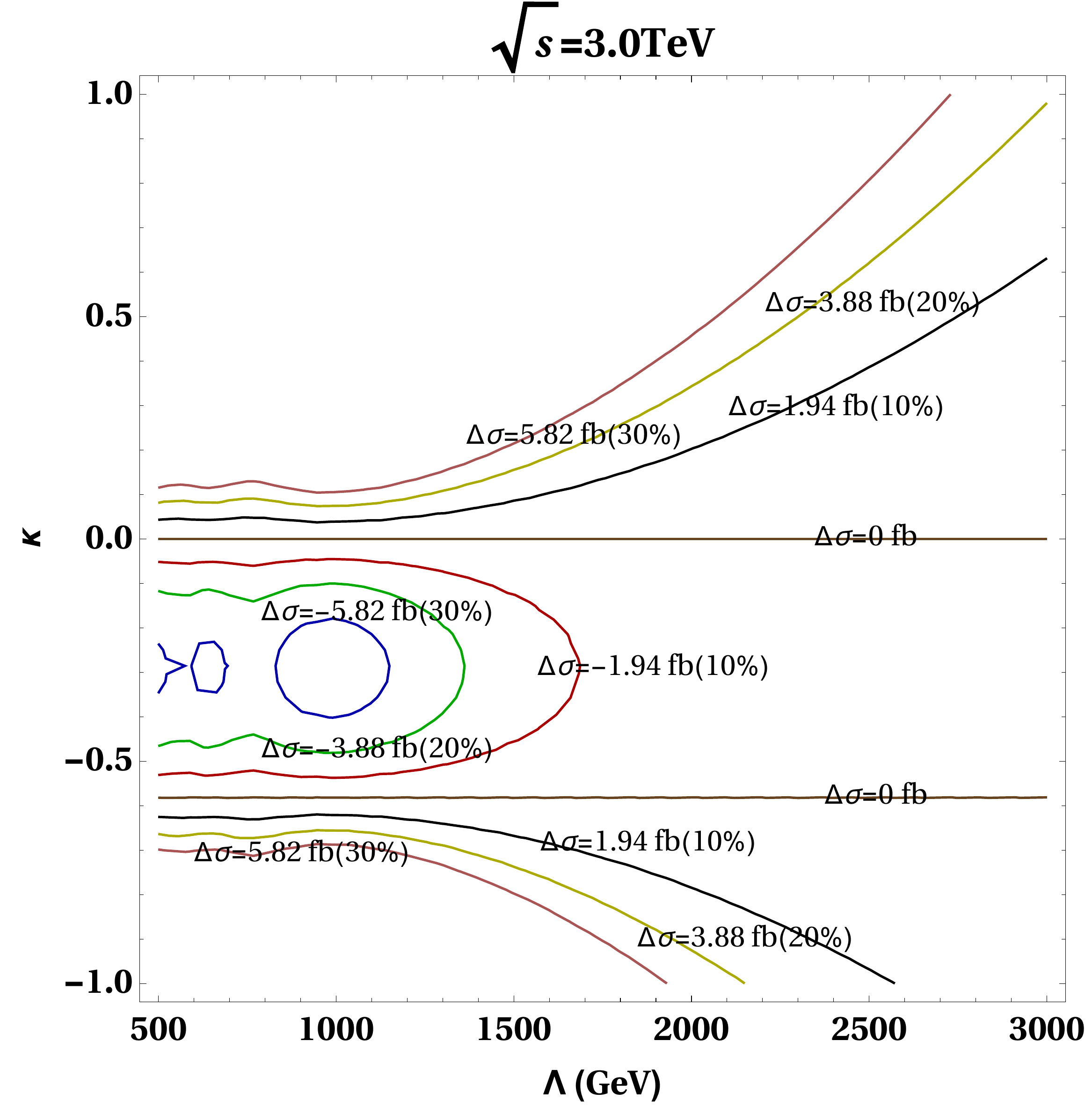}
\caption{\label{fig:2} The figure depicts the correction in NC cross-section ($\Delta \sigma$ in fb) with respect to the standard model ($\sigma_{SM}^{1.4}=87.833\,fb,\,\sigma_{SM}^{3.0}=19.402\,fb$) in the 
presence of non-vanishing covariant $\theta$-exact coupling $\kappa$ at CLIC for $\sqrt{s}=1.4 $ TeV (Left plot) and for $\sqrt{s}=3.0 $ TeV (Right plot) respectively.
Also shown the percentage of corresponding correction with respect to SM within the brackets.}
\end{figure}

The $\Theta$-exact covariant non-minimal coupling $\kappa$ appears linearly in the differential cross section expressions from 
equations \ref{diffphoton}-\ref{diffphotonZ}, while the NC scale ($\Lambda$) arises in the argument of the oscillatory function. 
In fact, there is no correlation between the non-minimal coupling $\kappa$ and the NC scale 
$\Lambda$. To see this, let us 
consider the photon mediated process,the differential cross section of which is given 
in equation \ref{diffphoton}. It has the term $4\,\kappa(\kappa+Q_{t})$. Notice that, the 
NC contribution will vanish either for $\kappa=0$ or $\kappa=-Q_{t}$ for all values of 
$\Lambda$. 
In the $V-A$ vertex, direct and interference terms of the differential cross-section, 
the coupling $\kappa$ appears as a linear function. To analyze the noncommutative effect, we next
construct the NC correction observable  
$$ \Delta \sigma = \sigma^{NC} - \sigma^{SM}$$
In Fig.\ref{fig:2}, we have made the contour plot in the plane of $\kappa$ and 
$\Lambda$ corresponding to different $\Delta \sigma$ values which ranges from $10\%$ to 
$30\%$ correction with respect to 
SM cross section corresponding to the machine energy $\sqrt{s}=1.4$ TeV and $\sqrt{s}=3.0$ TeV, respectively.
We see that the lower bound on $\Lambda$ increases with $\kappa$ as long as 
$\kappa$ is positive. From the plot, we find that the NC contribution vanishes at 
 $\kappa=0$ and $\kappa\simeq-0.596$. This behaviour can be understood from the photon mediated 
 process whose contribution is the predominant one. Here, specific domain of negative $\kappa$, 
 exhibits the negative residual NC effects ($\Delta \sigma=-ve$), null effects i.e. $\Delta \sigma=0$ 
 and positive residual NC effects ($\Delta \sigma=+ve$) as shown in Fig.\ref{fig:2} for distinctive 
 range of $\Lambda$ corresponding to the machine energy $\sqrt{s}=1.4$ TeV and $\sqrt{s}=3.0$ TeV, 
 respectively. The negative residual cross section tells that the SM cross section is greater 
 than the NC cross section.
So we can classify the $\kappa $ parameter region as the region where the 
coupling ($\kappa$) is positive, the residual cross section is positive and the region where 
the coupling ($\kappa$) is negative, the residual cross section can be either negative, zero or positive.
The negative residual effect, bounded by the region $\kappa \in [0,-0.596]$, gives rise a lower 
bound on the NC scale $\Lambda$.
In the negative $\kappa$ region, as $\kappa$ increases from $0$ to $\kappa_{max}$, for a given $\Delta \sigma$, the lower bound on 
$\Lambda$ increases, and it then start decreases as $\kappa$ varies from $\kappa_{max}$ 
up to $-0.596$ it decreases the lower 
bound on $\Lambda$. For $\kappa_{max}=-0.298$, one finds the lower bound on NC scale as 
$\Lambda \geq 780(1680)$ GeV (at the machine energy $\sqrt{s}=1.4(3.0)$TeV) corresponding to  
$\Delta \sigma \sim  10\%$. On the other hand the positive residual effects are monotonically 
increasing (decreasing) effects in the region where $\kappa \geq 0$ ($\kappa \leq -0.596 $).
\subsection{Optimal collision energy}\label{oce}
\begin{figure}[h]
\centering 
\includegraphics[width=.5\textwidth]{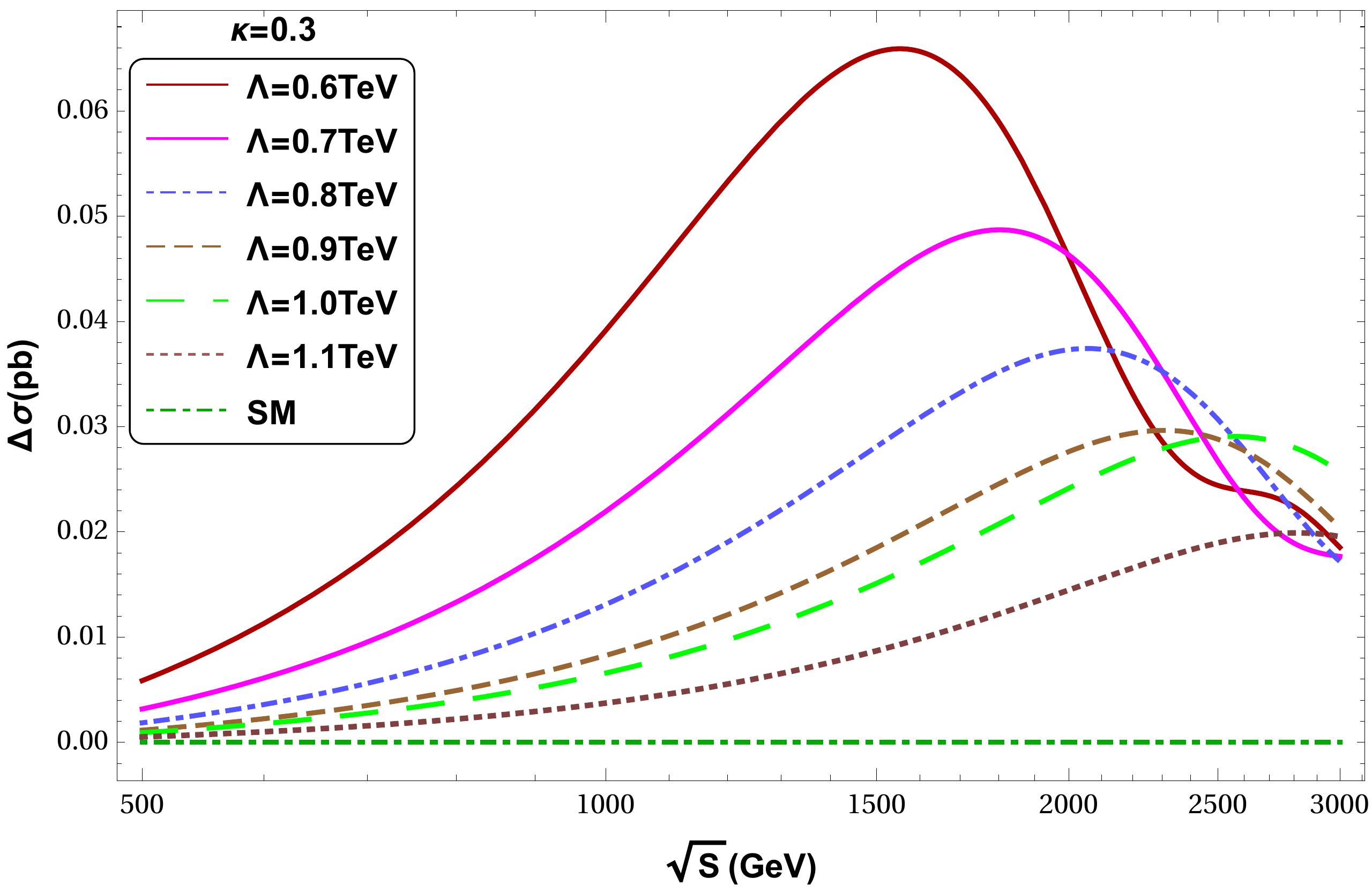}
\hfill
\includegraphics[width=.5\textwidth]{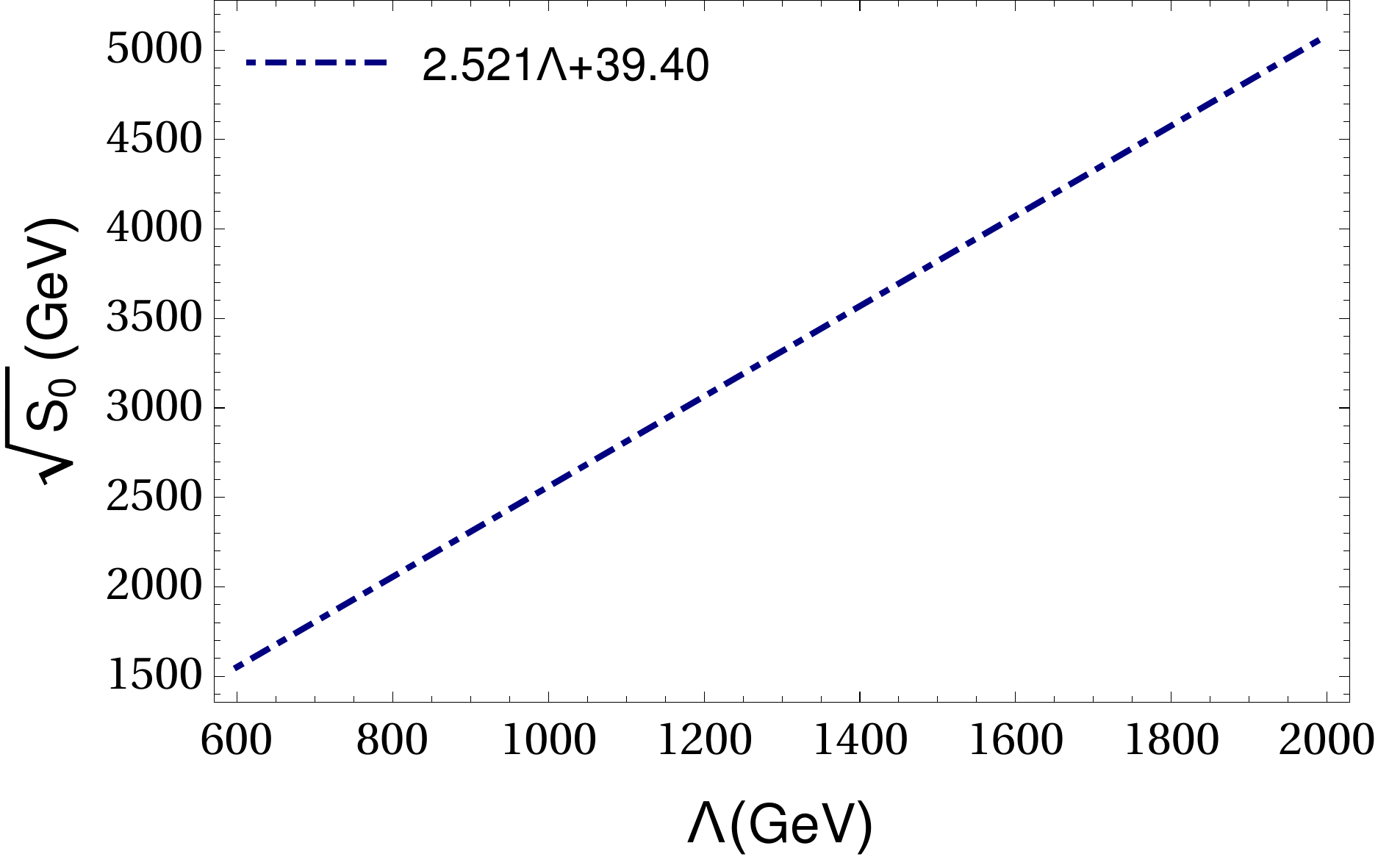}
\caption{\label{fig:3} The left figure shows that the difference between NC total cross section and SM cross section ($\Delta\sigma=\sigma_{NCSM}-\sigma_{SM} $) as of a function of machine energy $\sqrt{s}$ and
right figure shows probe of NC effects at optimum collision energy respectively.}
\end{figure}

The positive residual region in the Fig.\ref{fig:2} illustrates the monotonic behaviour of the 
coupling $\kappa$ with respect to the NC scale $\Lambda$, but it does not pinpoint where
 one can look for the NC signature. In contrast, it is not easier to locate the NC signature by arbitrarily scanning over the wide range of the collision energy.
 In Fig.\ref{fig:3} (Left plot), the positive Kurtosis distribution exhibits that the NC contribution 
 in the top pair production cross section first increases with the machine energy and then starts 
 decreasing after a certain threshold machine energy for a specific value of $\Lambda$ at a fixed coupling 
 $\kappa$.   
It is true for all value of $\Lambda$ and $\kappa$. This machine threshold energy (also called the optimal machine energy)
may be quite useful to look for the signature of the spacetime noncommutativity.
 One can obtain the linear relation by connecting all optimal collision energy (say $\sqrt{s_{0}}$) for respective value of the NC scale $\Lambda$ for a given $\kappa$.
The optimal collision energy expressions for different $\kappa$ value can be written as 
\begin{eqnarray}
 \sqrt{s_{0}} & = & 2.52151 \, \Lambda + 39.067 \, \quad \, (\kappa = 0.1) \\
 \sqrt{s_{0}} & = & 2.52105 \, \Lambda + 39.443 \, \quad \, (\kappa = 0.3) \\
 \sqrt{s_{0}} & = & 2.52107 \, \Lambda + 39.428 \, \quad \, (\kappa = 0.5) 
\end{eqnarray}
Here the NC scale $\Lambda$ and the optimum collision energy $\sqrt{s_{0}}$ are in units of GeV.
Notice that the optimal collision energy abruptly follows $2.5$ times (approx) the NC scale $\Lambda$.
The Fig.\ref{fig:3} (Right plot) shows that the allowed region (below line) of optimum lower limit of the NC scale which is obtained by optimum condition. 
These optimum region are not necessarily lower limit of the NC scale because the amount of NC cross section correction ($\bigtriangleup \sigma $) is not taken into account.
Further, Fig.\ref{fig:3} (Left plot) attains optimum point at certain value of machine energy ($\sqrt{s}$) for each NC scale which is known as optimum lower limit. 
This is the evident of the $TeV$ scale spacetime noncommutativity which is residing under the line $ 2.5 \, \Lambda$ (approx).
The positive value of $\kappa$ doesn't affect the optimal collision energy relation
but it plays an important role in the magnitude of the excess total cross section (see Fig.\ref{fig:2}). 
We see that one requires the collision energy to be greater than the NC scale in order to probe 
the spacetime noncommutativity at the future collider.


\subsection{Inferring $ \kappa$ in the light of azimuthal anisotropy}\label{sec:phiasy}
The azimuthal angular distribution of the top quarks pair (after reconstructing the $p_{T}$ of the top quark) can be an useful tool to probe the 
beyond the standard model (BSM) physics. Since $\Theta^{0i}\neq 0$ in our $\Theta^{\mu \nu}$ structure which is invariant
under VSR Lorentz group symmetry, it makes NC theory rotationally non invariant. 
\begin{figure}[h]
\centering 
\includegraphics[width=.7\textwidth]{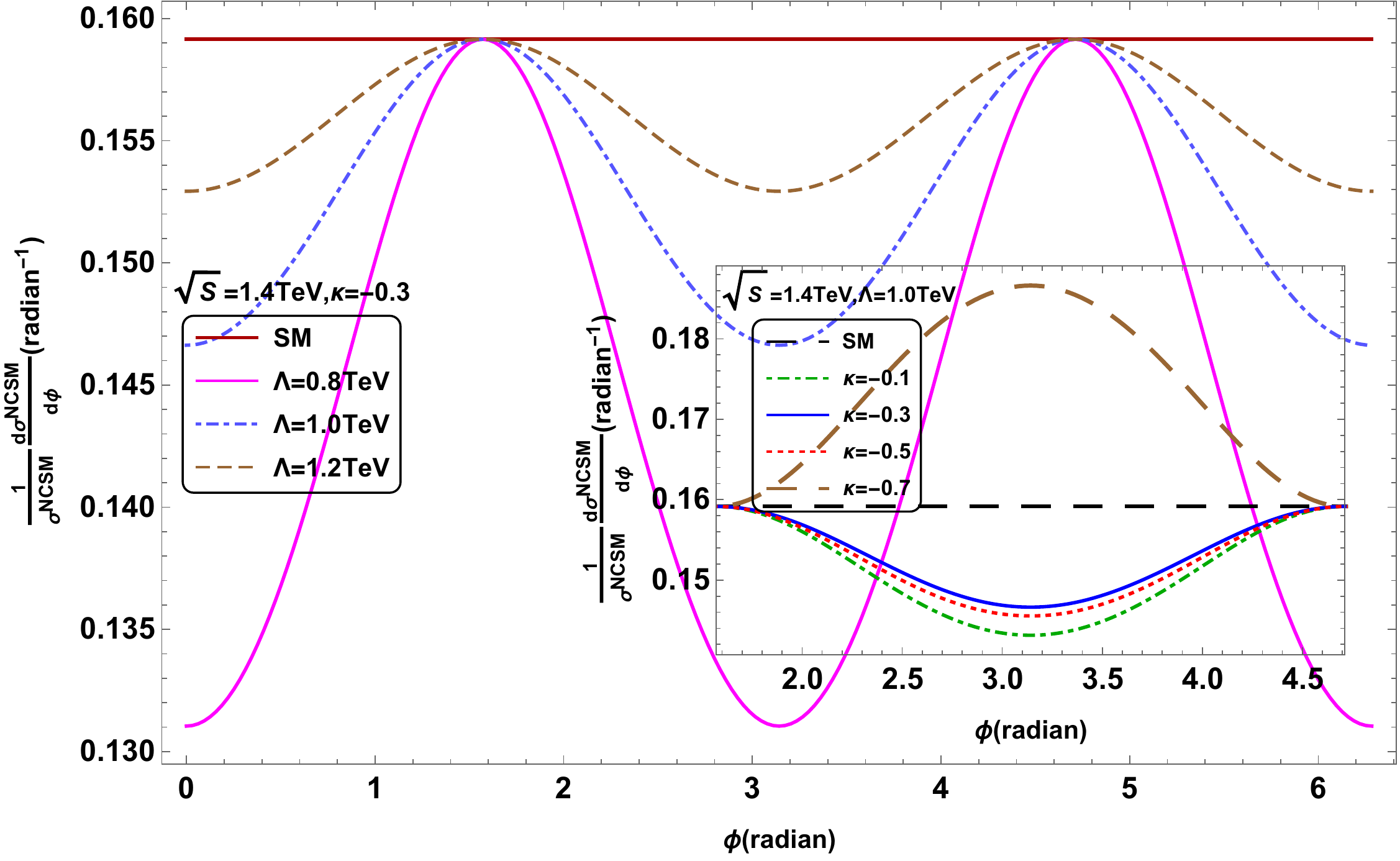}
\caption{\label{fig:4} The figure shows normalized azimuthal anisotropy when $\kappa =-0.3$ at $\sqrt{s}=1.4$TeV. Inset shows that the normalized azimuthal anisotropy for $\kappa<0$ 
and the azimuthal angle $\phi$ taken between $\frac{\pi}{2}$ and $3\frac{\pi}{2}$.}
\end{figure}
Here the non zero $\Theta^{0i}$ produces the azimuthal anisotropy through momentum weighted $\Theta$ product which is $p_{4}\Theta p_{3}$. 
The amplitude of the anisotropy is gradually decreases by $f(1/\Lambda^{2})$. Depending up on the values of $\kappa$, 
the normalized azimuthal distribution has two different behaviour which is $\cos^{2}(f(1/\Lambda^{2}) \cos[\phi])$ for $0 >\kappa > -0.596$ in Fig.\ref{fig:4} 
and $\sin^{2}(f(1/\Lambda^{2}) \cos[\phi])$ for $\kappa > 0$ and $\kappa < -0.596$ in Fig.\ref{fig:5}.
\begin{figure}[h]
\centering 
\includegraphics[width=.7\textwidth]{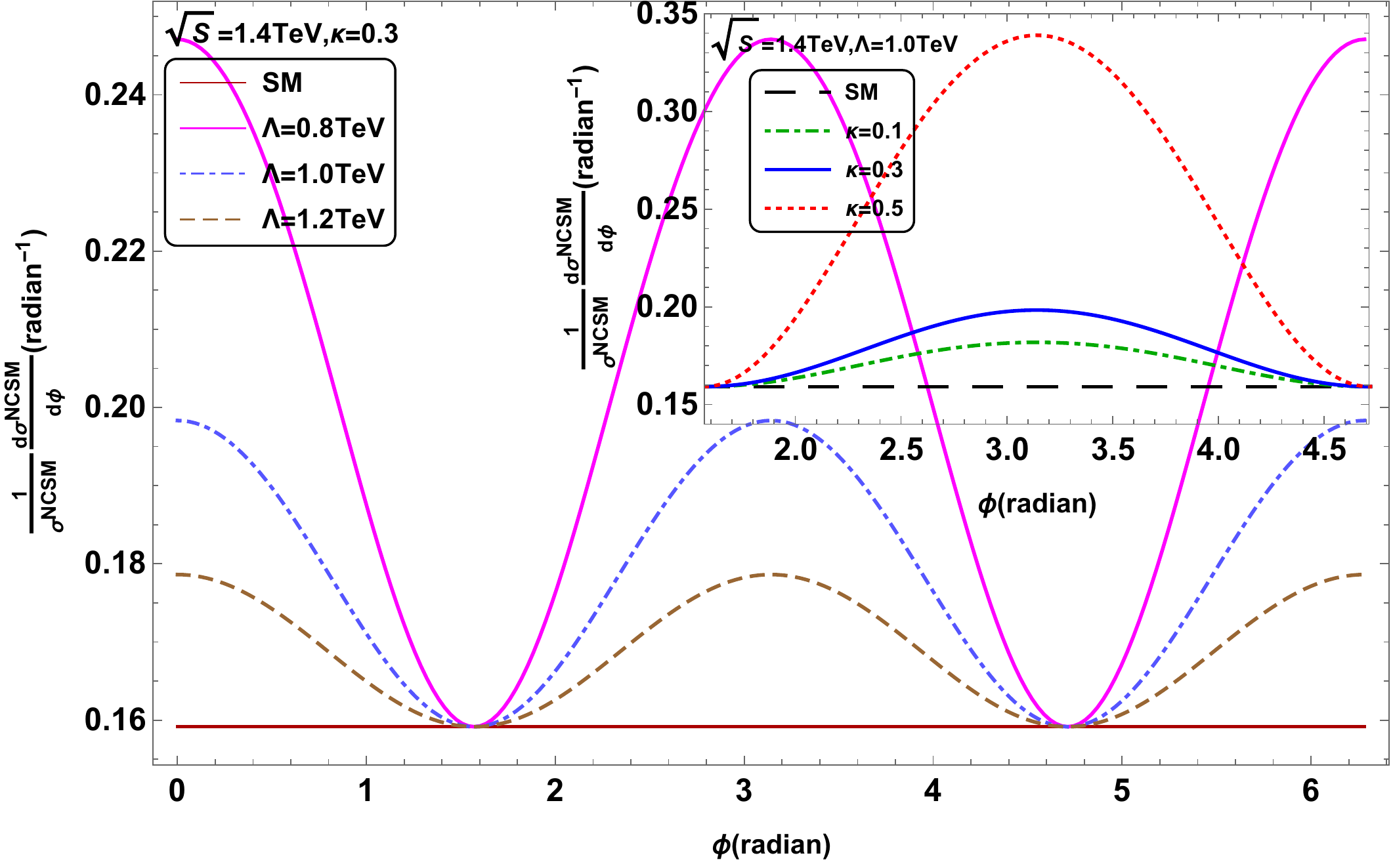}
\caption{\label{fig:5} Same as Fig.\ref{fig:4} for $\kappa =0.3$ at $\sqrt{s}=1.4$TeV. Inset shows that the normalized azimuthal anisotropy for $\kappa>0$ 
and the azimuthal angle $\phi$ taken between $\frac{\pi}{2}$ and $3\frac{\pi}{2}$.  }
\end{figure}
This behaviour dominantly coming from equation \ref{diffphoton} in the domain $0 >\kappa > -0.596$ and $\kappa > 0$ and $\kappa < -0.596$ respectively.
However, the prediction of the signature of spacetime noncommutativity in these two domain are quite different. The domain
$\kappa>0$ has discussed in the above Sec.\ref{oce} and the allowed region of the optimum lower bound on $\Lambda$ are shown in Fig.\ref{fig:3} (Right plot).
The statistical analysis is the prominent way to find the signature of the spacetime noncommutativity which can be considered in the domain where $0 >\kappa > -0.596$.
So one can choose the azimuthal anisotropy as an observable to do statistical analysis namely $\chi^{2}$ analysis. 
The azimuthal asymmetry would give the fluctuated (excess or less) number of events around standard model flat value from $\phi=0$ to $\phi=2\pi$.
 \begin{figure}[h]
 \centering 
 \includegraphics[width=.45\textwidth]{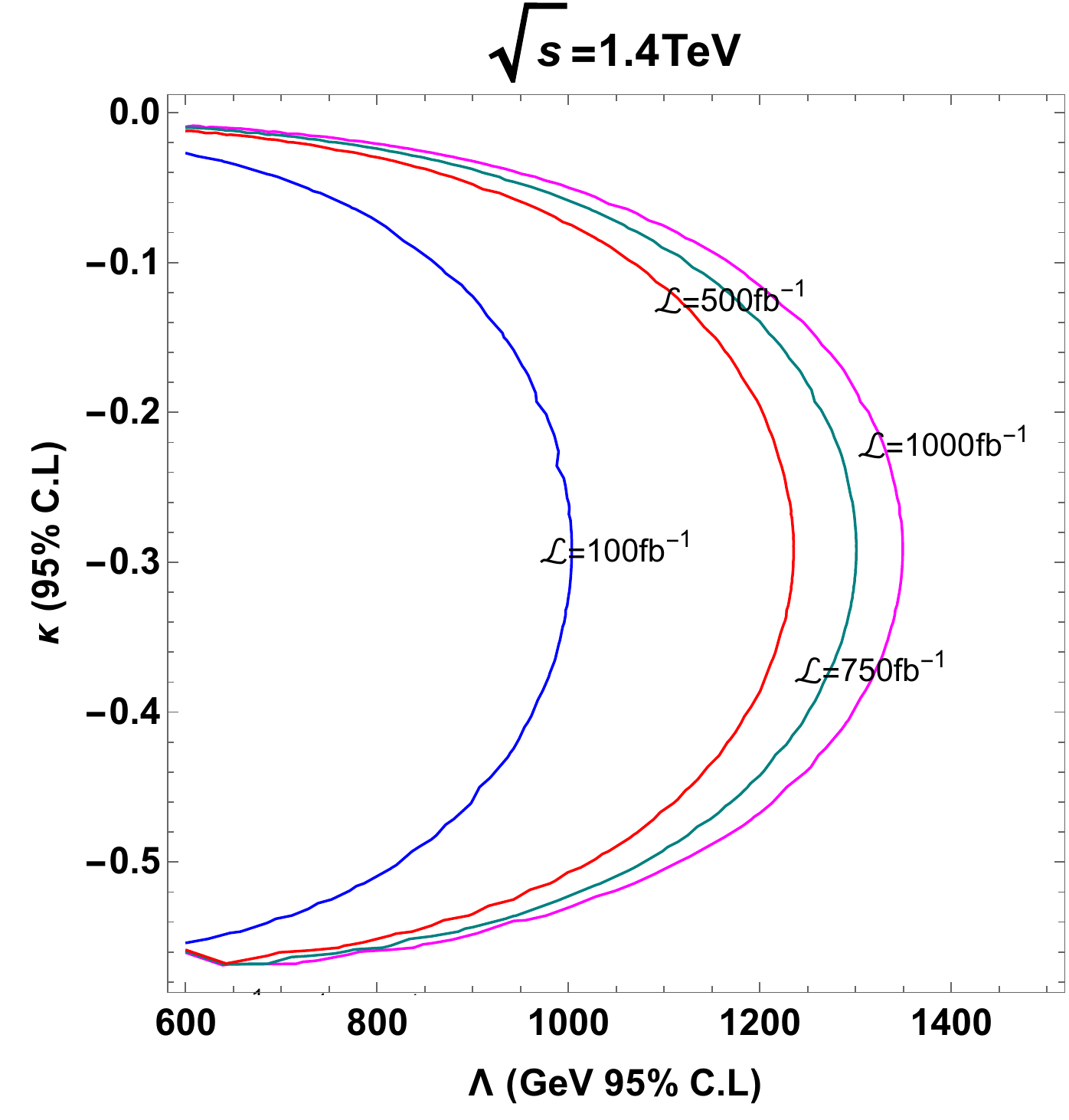}
  \hfill
 \includegraphics[width=.45\textwidth]{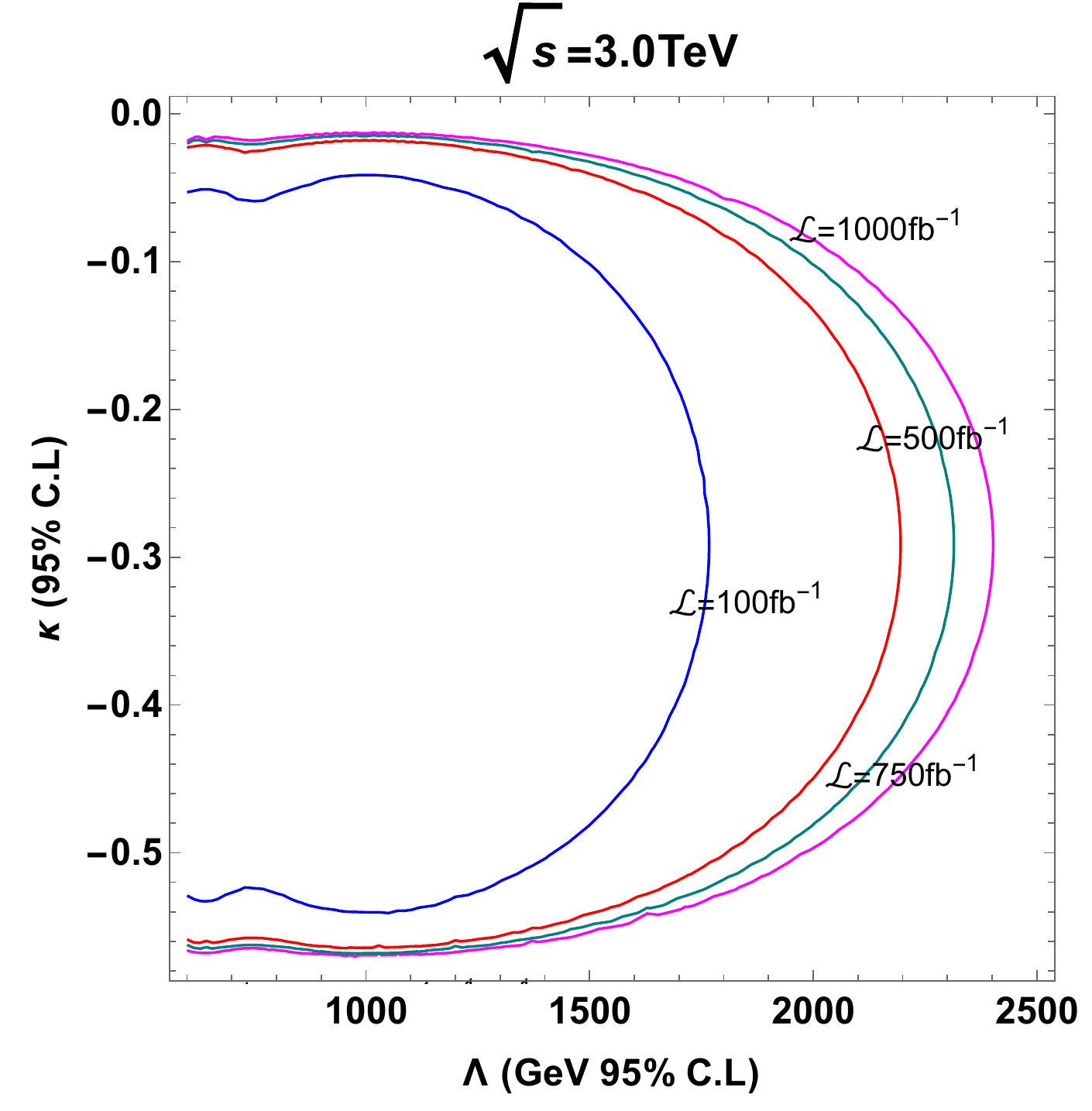}
 \caption{\label{fig:6}
 The left (right) figure depicts $\chi^{2}$ statistical test ($ 95\%  C.L$) of the noncommutative signal event
 arises due to azimuthal anisotropy at $\sqrt{s}=1.4$ TeV ($\sqrt{s}=3.0$ TeV).}
 \end{figure}
Here we have taken twelve bins, each has $\phi=\pi/6$ and totally eleven d.o.f in the $\chi^{2}$ value. The $\chi^{2}$ has defined as
$$ \chi^{2}(\Lambda) = \sum_{i=1}^{12} \frac{(N_{i}^{NCSM}(\Lambda) - N_{i}^{SM})^{2}}{N_{i}^{SM}}; \quad \, \quad N_{i}=\mathcal{L} \int\frac{d \sigma_{i}}{d \phi} d\phi$$
The Fig.\ref{fig:6} depicts the statistical value of $\chi^{2}_{0.050}=19.675$ for both $\sqrt{s}=1.4$ TeV and $\sqrt{s}=3.0$ TeV for given integrated
luminosity $\mathcal{L}$. Thus the luminosity contours excludes the lower bound at $95\%$ confidence level in the $\kappa - \Lambda$ plane. 
\begin{table}[h]
\centering
\begin{tabular}{|l|c|c|c|c|}
\hline	
Integrated luminosity ($\mathcal{L}$) & $100\, fb^{-1}$ & $500\, fb^{-1}$ & $750\, fb^{-1}$ & $1000\, fb^{-1}$\\
\hline 
  & & & & \\
Lower limit on $\Lambda$: $\sqrt{s}=1.4$TeV  & $1.011\, TeV$  & $1.235\, TeV$ & $1.306\, TeV$ & $1.352\, TeV$ \\ 
& & & & \\
Lower limit on $\Lambda$: $\sqrt{s}=3.0$TeV  &  $1.775\, TeV$ & $2.20\, TeV$  & $2.318\, TeV$ & $2.406\, TeV$ \\ 
\hline
\end{tabular}
\caption{\label{tab:1} The lower bound on noncommutative scale $\Lambda$ ($95\%$C.L) obtained by $\chi^{2}$ analysis at $\kappa_{max}=-0.296$ for four different integrated luminosity.}
\end{table}
The lower limit on the NC scale $\Lambda$ ($95\%$C.L) at $\kappa_{max}=-0.296$ is given below in table.\ref{tab:1}.
Here one can notice that the noncommutative scale is sensitive to the integrated luminosity ($\mathcal{L}$) which is more
clearly given in the Fig.\ref{fig:13}. The slope of the luminosity curve in the Fig.\ref{fig:13} (red curve) is smaller in the higher luminosity region
than the lower luminosity region. Such type of luminosity sensitivity has noticed in \cite{Zahra,PTEP}. In \cite{PTEP},
authors considered profile likelihood ratio for two hypothesis test which is \textit{signal$+$background} and \textit{background}. We consider the 
background events as a SM $t \overline{t}$ pair events, thus our analysis simply boils down in to $\chi^{2}$. So one can compare the behaviour of  
$\Lambda$ from those two different approaches which are quite same results though they are different final production states too.

\section{Helicity amplitude techniques in $ \Theta$-exact NCSM}\label{sec:calcul2}
\begin{figure}[h]
\centering 
\includegraphics[width=.75\textwidth]{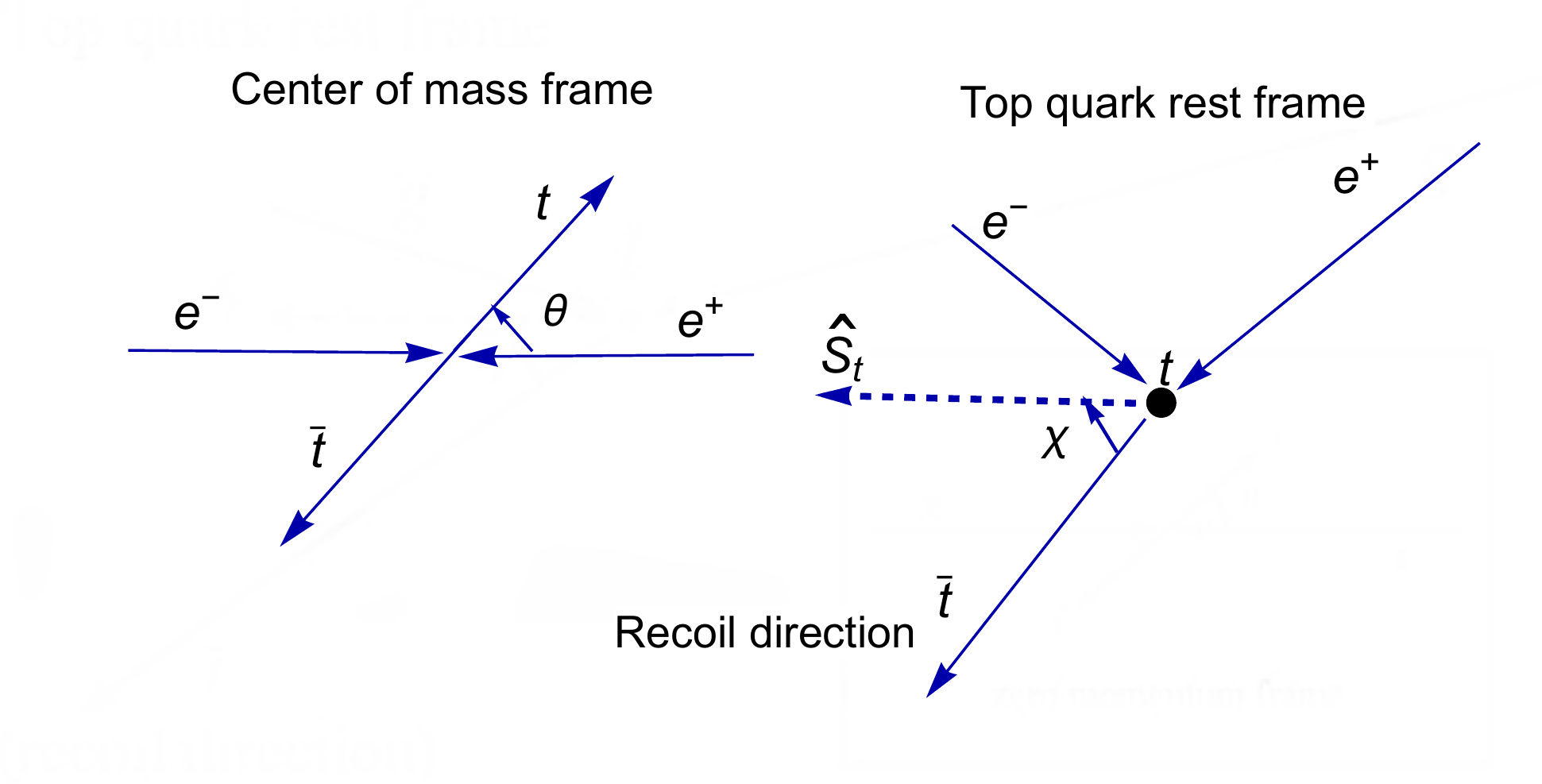}
\caption{\label{fig:7} Top and anti-top production in the COM frame and top spin vector in the top rest frame.}
\end{figure}

The study of polarization allows us to probe the chirality of the interactions between the top quark and gauge bosons 
in the SM as well as NCSM. The top polarization can be clearly analyzed in the $e^{-}e^{+}$ collider from $t \, \overline{t}$ pair production.
We define the spin of the top quark and anti-top quark as shown in the Fig.\ref{fig:7} at the production plane, 
so the transverse momentum of the top quark is zero and the spin four-vectors are back to back in the center of mass frame.
Here $\widehat{s}$ is the spin vector of the top quark which makes an angle $\chi$ with anti-top quark
in the recoil direction. In the special case, we take the spin angle $\chi=0$ or $\pi$, which tells that the spin direction is
along the direction of the top quark momentum and it defines the helicity states of the top quark in the $t\,\overline{t}$ rest frame. If the massive top quark 
four momentum is $P^{\mu}=(E;\, \overrightarrow{\textbf{p}})$ then the spin four-vector defined as\cite{Haber,HelicLC}
\begin{equation}
 s^{\mu} = (2\lambda) \frac{1}{m}(|\overrightarrow{\textbf{p}}|;\, E \, \widehat{\textbf{p}})
\end{equation}
Here $(2\lambda)=\pm1$ is the twice of the spin $1/2$ particle helicity ($\lambda$) and the spin four-vector satisfies 
$ s.p=0 $ and $s.s=-1$. One can derive the helicity projection operators as follows
\begin{eqnarray}
 u(p,\lambda)\, \overline{u}(p,\lambda) & = & \frac{1}{2}(1+\gamma_{5} \slashed{s})(\slashed{p}+m) \\
 v(p,\lambda)\, \overline{v}(p,\lambda) & = & \frac{1}{2}(1+\gamma_{5} \slashed{s})(\slashed{p}-m) 
\end{eqnarray}
Here we works in the commutative limit of the spacetime noncommutativity which does not change the Lorentz structure of the top quark spin. 
The generic differential cross section in the helicity amplitude technique at the center of mass frame (CM) for 
$a(\lambda_{a}) + b(\lambda_{b}) \longrightarrow c(\lambda_{c}) + d(\lambda_{d})$ is 
$$ \frac{d\sigma}{d\Omega_{CM}}=\frac{1}{64 \pi \, s} \left(\frac{p_{f}}{p_{i}}\right)\sum _{\lambda_{a}\lambda_{b}\lambda_{c}\lambda_{d}}|\mathcal{M}_{\lambda_{a}\lambda_{b};\lambda_{c}\lambda_{d}}|^{2}$$
Here $p_{i}(p_{f})$ is the initial (final) CM momentum, $\sqrt{s}$ is the 
CM-energy and $d\Omega_{CM}=d\cos\theta d\phi$. The detailed expressions for matrix element squared of all non vanishing helicity states are given in the appendix \ref{app:helincxsec}. 
\subsection{Helicity correlation and top quark left-right asymmetry}\label{sec:hclrttbar}
\textbf{Helicity correlation} \\
The top quark is the heavier quark in the standard model. It decays weakly into $b$ quark and $W$ boson before hadranization process happens.
In fact the lifetime ($\tau_{t}\approx 5\times 10^{-25}$s) of the top quark is lesser than hadranization time ($\tau_{QCD}\approx 1/\Lambda_{QCD}=(200\,MeV)^{-1}\approx 3.3\times 10^{-24}$s)\cite{Peskin,Bigi,spincorr} 
as well as spin-decorrelation time ($\tau_{spin}\approx m_{t}/\Lambda^{2}_{QCD} \approx 3\times 10^{-21}$s) from spin-spin
interactions with the light quarks generated in the fragmentation process\cite{spincorrbook,D0}.
\begin{figure}[h]
\centering 
\includegraphics[width=.5\textwidth]{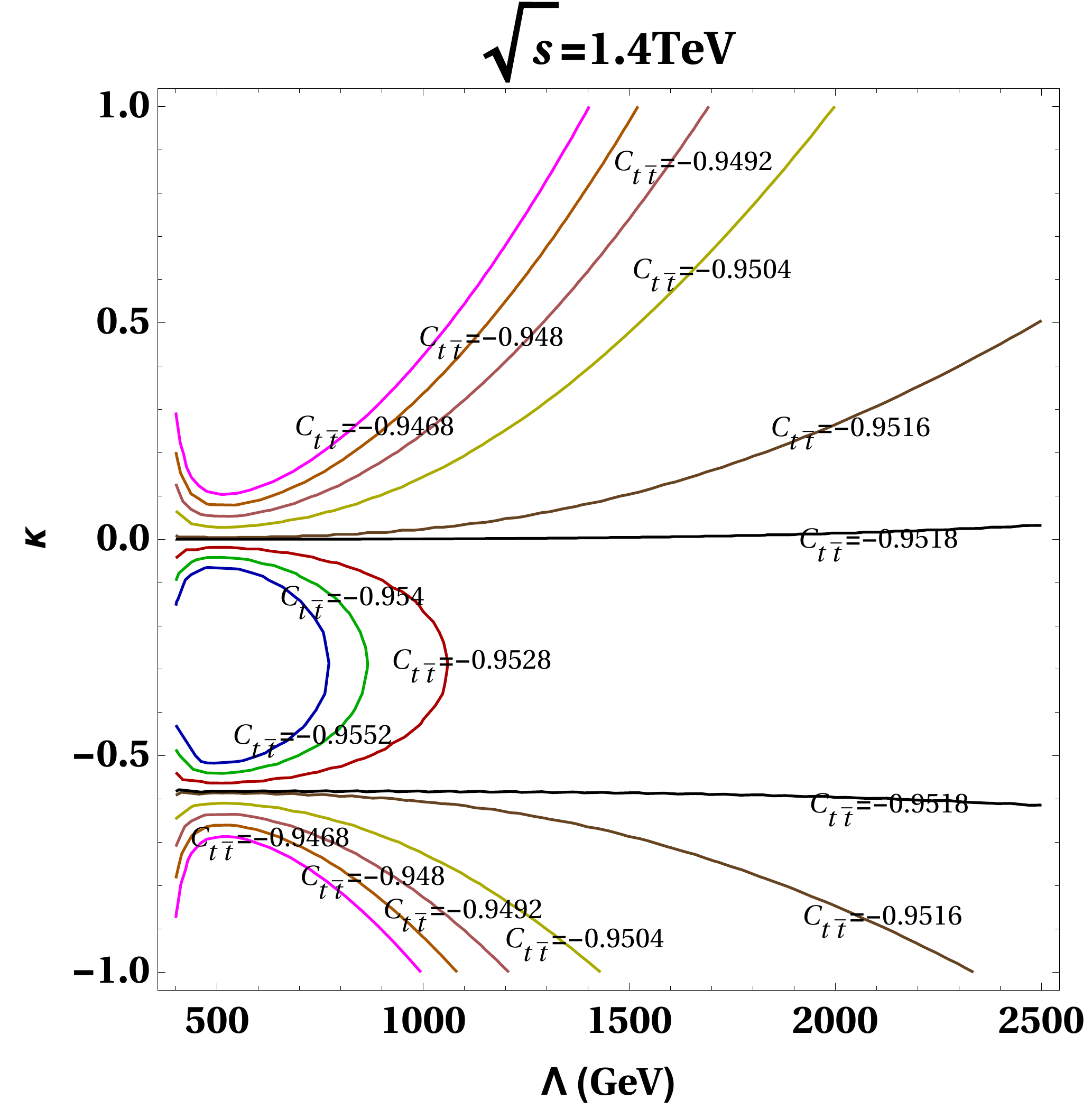}
\hfill
\includegraphics[width=.5\textwidth]{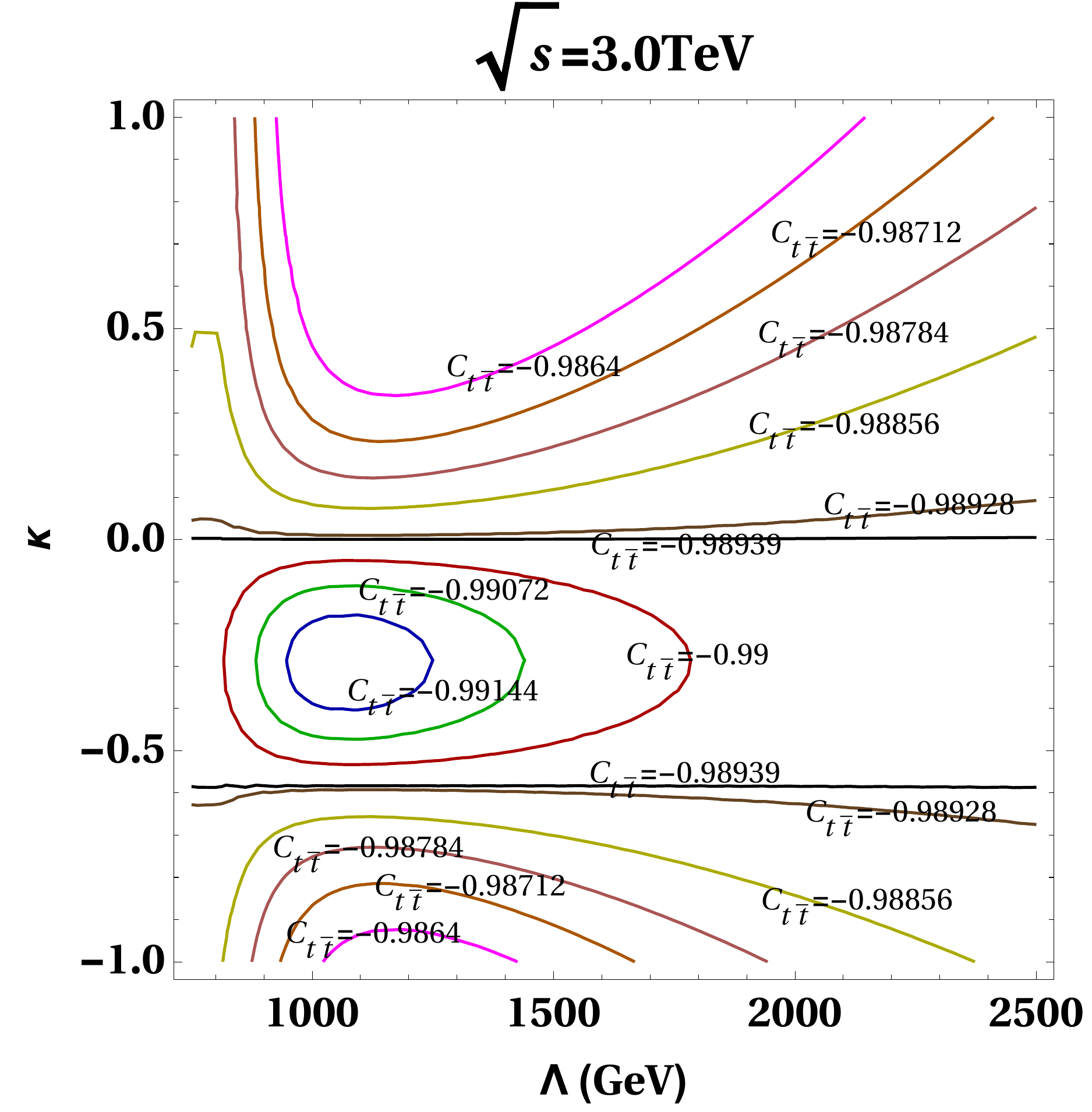}
\caption{\label{fig:8} Helicity correlation between the top and anti-top quark.}
\end{figure}
If the top quark spins are correlated when they are produced as a pair then their decay products are 
correlated with their spin and hence the decay products of the top quarks are correlated by naturally. 
The $t\,\overline{t}$ spin correlation reported in the CDF II detector \cite{CDFII,D0} by reconstructing lepton plus jets decay channel from $p\,\overline{p}$ collisions.
The measurement agrees with the SM QCD calculation by adding gluon fusion process. But the decay width of the top quark has deviation from SM calculation\cite{CDF}.
In the case of spacetime noncommutativity, the top decay were studied in the minimal $\Theta$-expanded NCSM\cite{namit,najafa,Zahra2}. \\
However this non-minimal $\Theta$-exact SW map approach will not participate in the tree level top decay, it appears only in the loop level.
Since there is no right-handed $W$ boson in the NCSM and we extended non-minimal star product among abelian gauge field and 
SM fermions with satisfying the gauge transformation which is given in equation.\ref{gaugetrans}, one cannot have a non-minimal coupling $\kappa$ with top quark in the charged current.
Thus the effect of spacetime noncommutativity on spin correlation can be determined from the top quark pair production.

Considering the top pair production, the measured spin correlation depends on the choice of the spin basis. There are three types of basis\cite{Parkelc}, which are i) beam line basis 
ii) helicity basis and iii) off-diagonal basis. Here we work in the helicity basis (which admits the center of mass frame in which the top
spin axis is defined) to find the top quark spin correlation.
 The helicity correlation factor defined as\cite{Parkelc,Parkelhc}
\begin{equation}
 C_{t\overline{t}}=\frac{\sigma_{LL}+\sigma_{RR}-\sigma_{LR}-\sigma_{RL}}{\sigma_{LL}+\sigma_{RR}+\sigma_{LR}+\sigma_{RL}}
\end{equation}
\begin{figure}[h]
\centering 
\includegraphics[width=.47\textwidth]{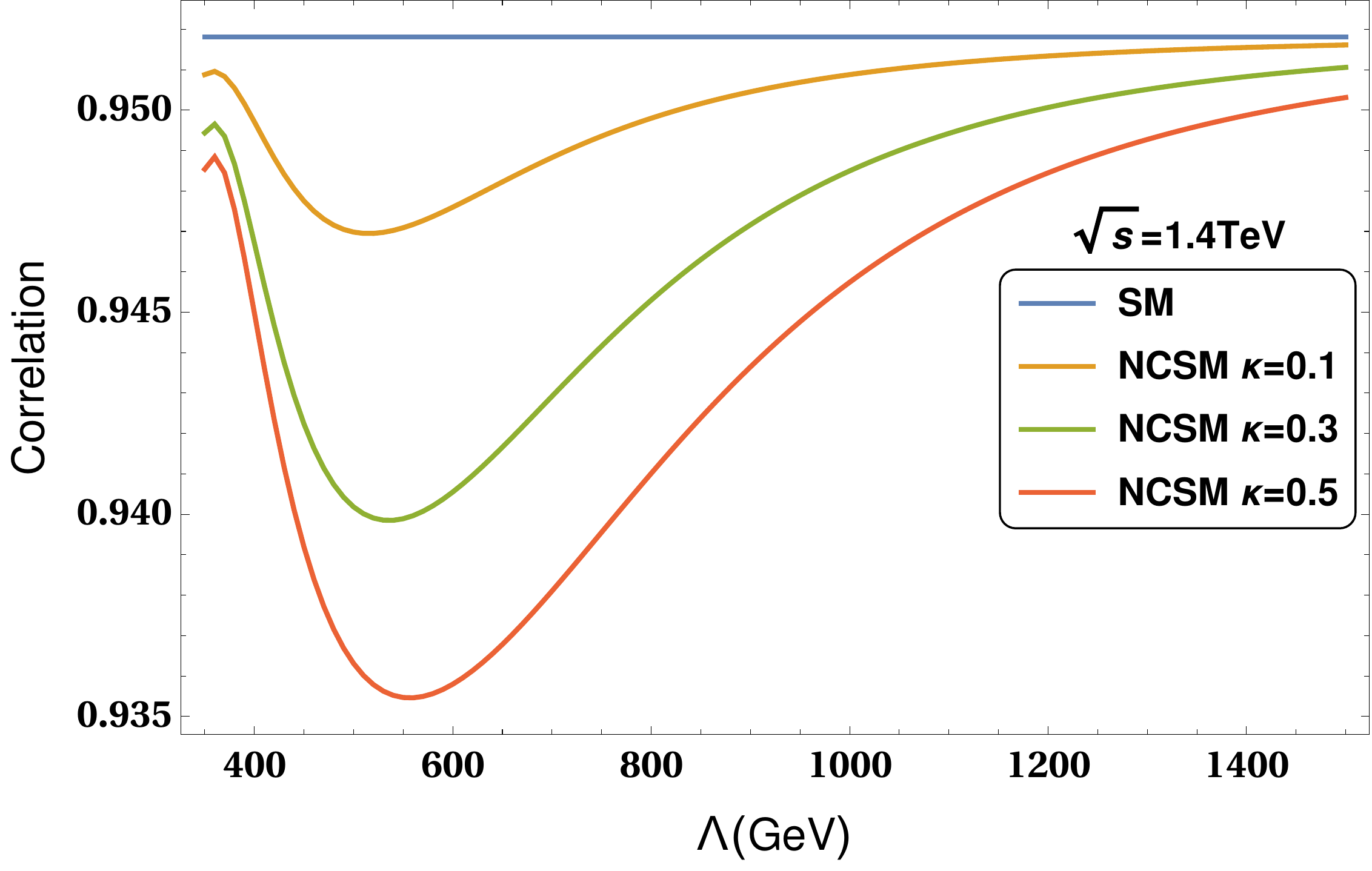}
\hfill 
\includegraphics[width=.47\textwidth]{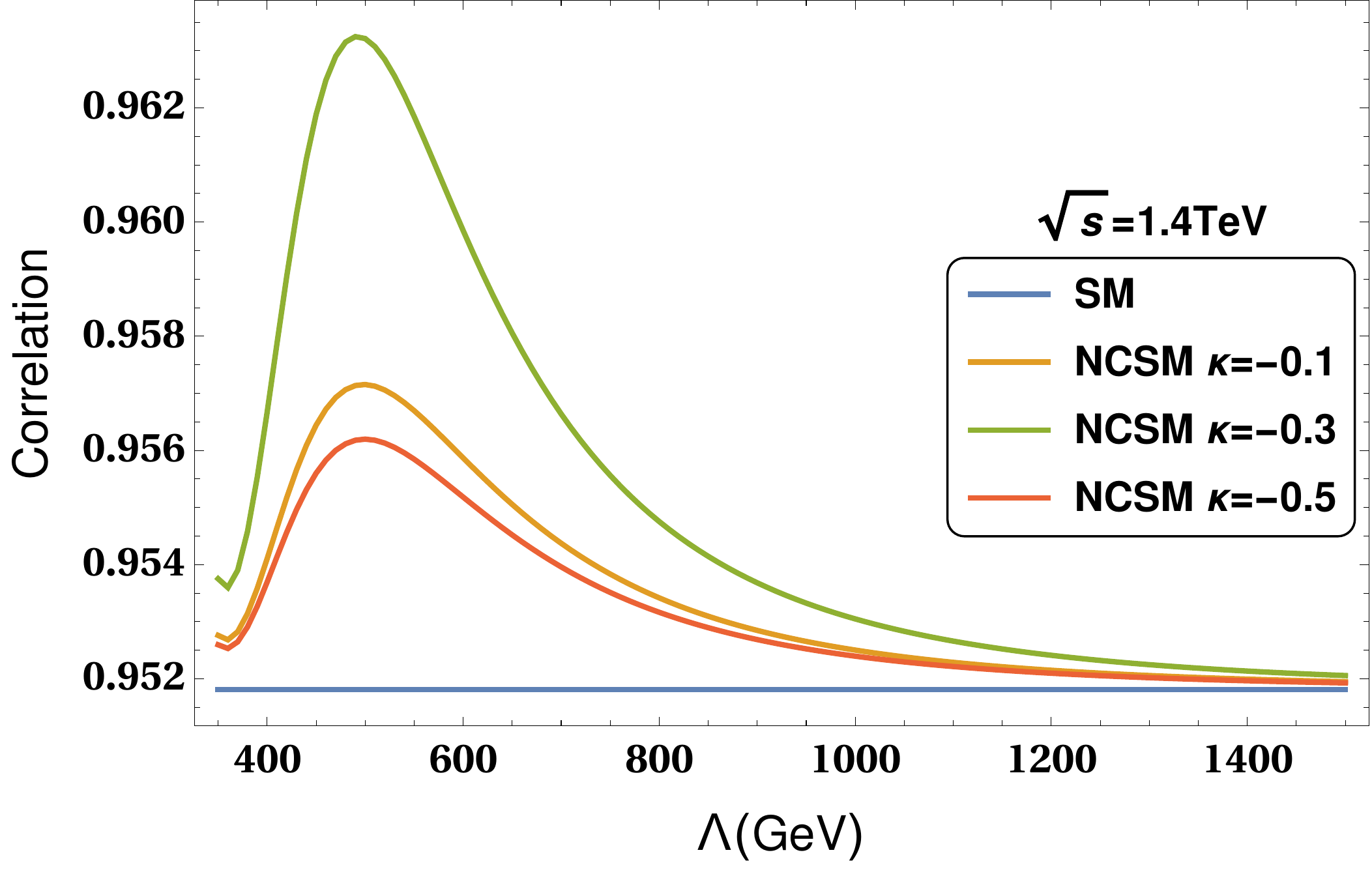} 
\caption{\label{fig:9} Helicity correlation ($|C_{t \overline{t}}|$) between the top and anti-top quark at $\sqrt{s}=1.4$ TeV in the presence of positive values of non-minimal coupling $\kappa$ (Left plot) and also negative (Right plot).}
\end{figure}
Here $\sigma_{i j} \quad (i,j=L,R) $ represents the total cross section of the final state top, anti-top helicity. Notice that $\sigma_{LL} \equiv \sigma_{RR}=0$ when 
$\beta=1$ i.e $\sqrt{s}>>m_{t}$ which is ultra-relativistic limit. In the ultra-relativistic limit, the 
helicity correlation is $C_{t\overline{t}}=-1$ when $LR$ configuration equals to $RL$ configuration in the $t \overline{t}$ pair production which are solely due to left-right helicity symmetry at high energy.

\begin{figure}[h]
\centering 
\includegraphics[width=.47\textwidth]{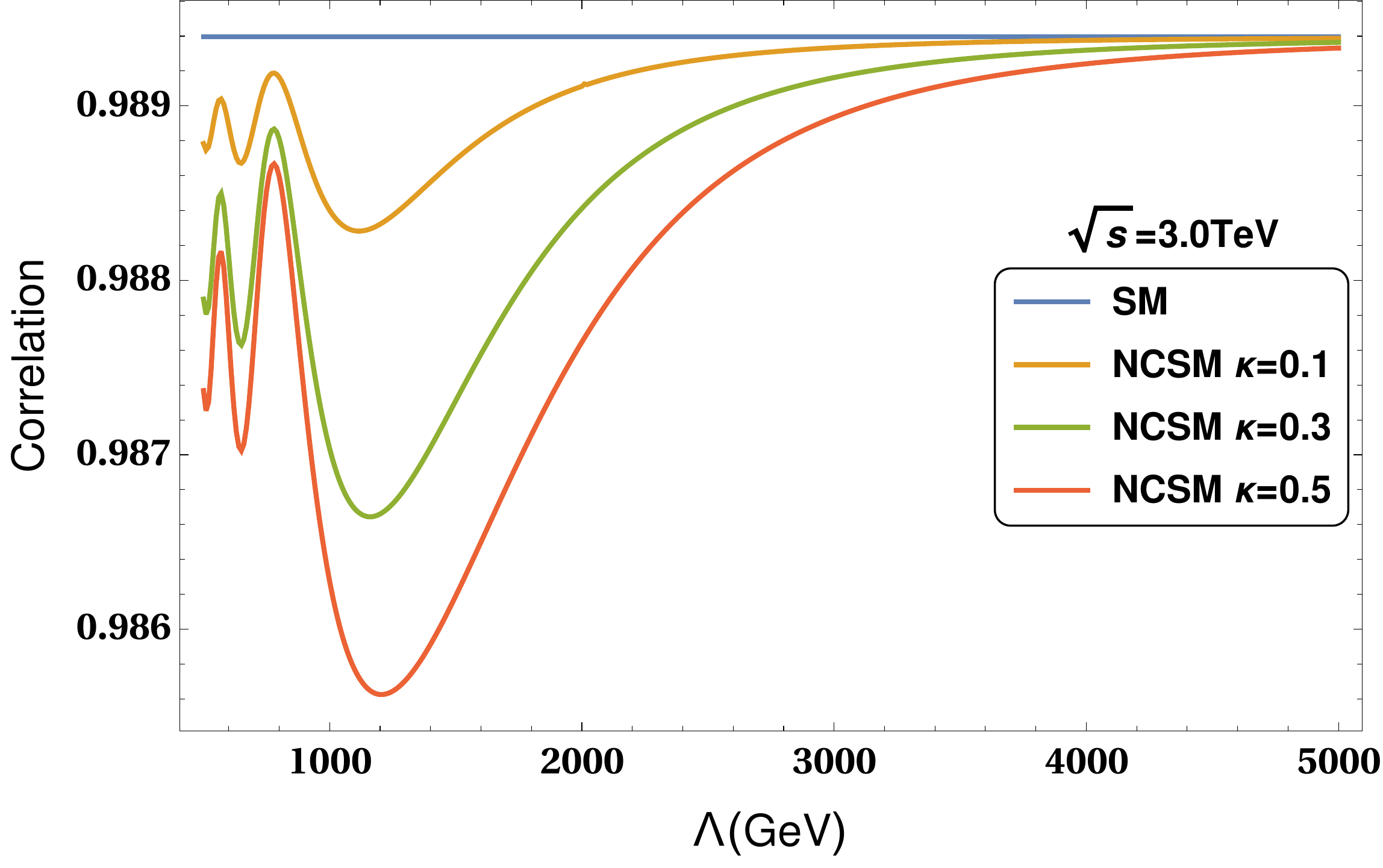}
\hfill
\includegraphics[width=.47\textwidth]{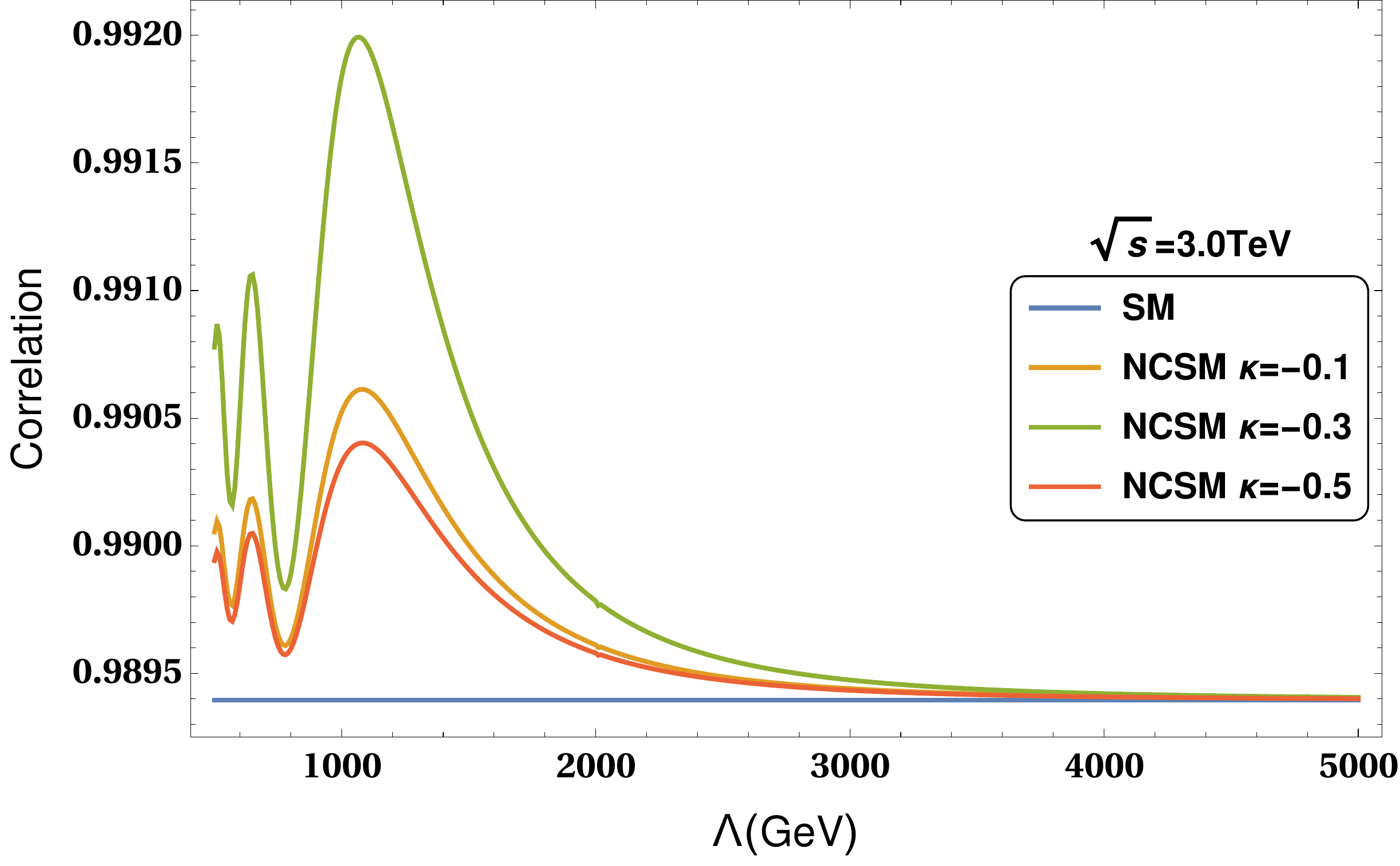}
\caption{\label{fig:10} Same as Fig.\ref{fig:9} for $\sqrt{s}=3.0$ TeV.}

\end{figure}
\noindent The top quark SM helicity correlation as shown in Fig.\ref{fig:8}, which are $C_{t \overline{t}}=-0.9518(-0.9894)$ 
at $\sqrt{s}=1.4(3.0)$ TeV  respectively. Here the negative sign emphasize that the opposite final state helicity cross section are usually dominant over the one with same helicity final
state cross section in the $e^{-}e^{+}$ colliders. In the region where $\kappa >0$ and $\kappa <-0.596$, the NC helicity
correlation is lesser than the SM helicity correlation for all values of NC scale $\Lambda$ and when $\Lambda \rightarrow \infty$ one recovers the SM results.
The another region $0 > \kappa > -0.596$, the value of $C_{t\overline{t}}$ reduces gradually when the NC scale increases.
But here the NC helicity correlation are greater than the SM correlation.

\noindent Notice that the region $\Lambda < 350$GeV at$\sqrt{s}=1.4$TeV and $\Lambda < 800$GeV at$\sqrt{s}=3.0$TeV are excluded because
which are arises due to unphysical oscillatory behaviour near lower values of NC scale. 
The top quark correlations for $\kappa >0$ and $\kappa < -0.596$ are given in Fig.\ref{fig:9} and Fig.\ref{fig:10} which shows that the 
optimal correlation values than SM value are located at $\Lambda \approx520$ GeV ($\sqrt{s}=1.4$ TeV) and $\Lambda \approx1150$ GeV ($\sqrt{s}=3.0$ TeV) respectively.  
Thus one can conclude that if any measured helicity correlation deviates from SM, our result will put a lower bound on NC scale 
which are $\Lambda \geq 520$GeV ($\sqrt{s}=1.4$TeV) and $\Lambda \geq 1150$GeV ($\sqrt{s}=3.0$TeV) respectively. 
The non-minimal coupling $\kappa$ can be arbitrary in the positive region which can be restricted by statistical significance 
of the experimental results but in the negative region it can be taken as $\kappa_{max}=-0.296$.\\

\noindent \textbf{Top quark left-right asymmetry}\\
In the top quark pair production, we can predict the dominant helicity of the top quark by calculating the top quark
left-right asymmetry. The polarized top quark decay produced $b$ quark and longitudinally polarized $W$ boson dominantly.
The ratio of the decay rate for polarized top quark and unpolarized top quark ($F_{0}=\left[\Gamma(t\rightarrow b\,W_{L})/ \Gamma^{total}(t\rightarrow b\,W)\right]$) 
was measured by CDF\cite{CDF} which is $F_{0}=0.91 \pm 0.37 \pm 0.13$.
The SM value is $F_{0}=0.701$. 
\begin{figure}[h]
\centering 
\includegraphics[width=.5\textwidth]{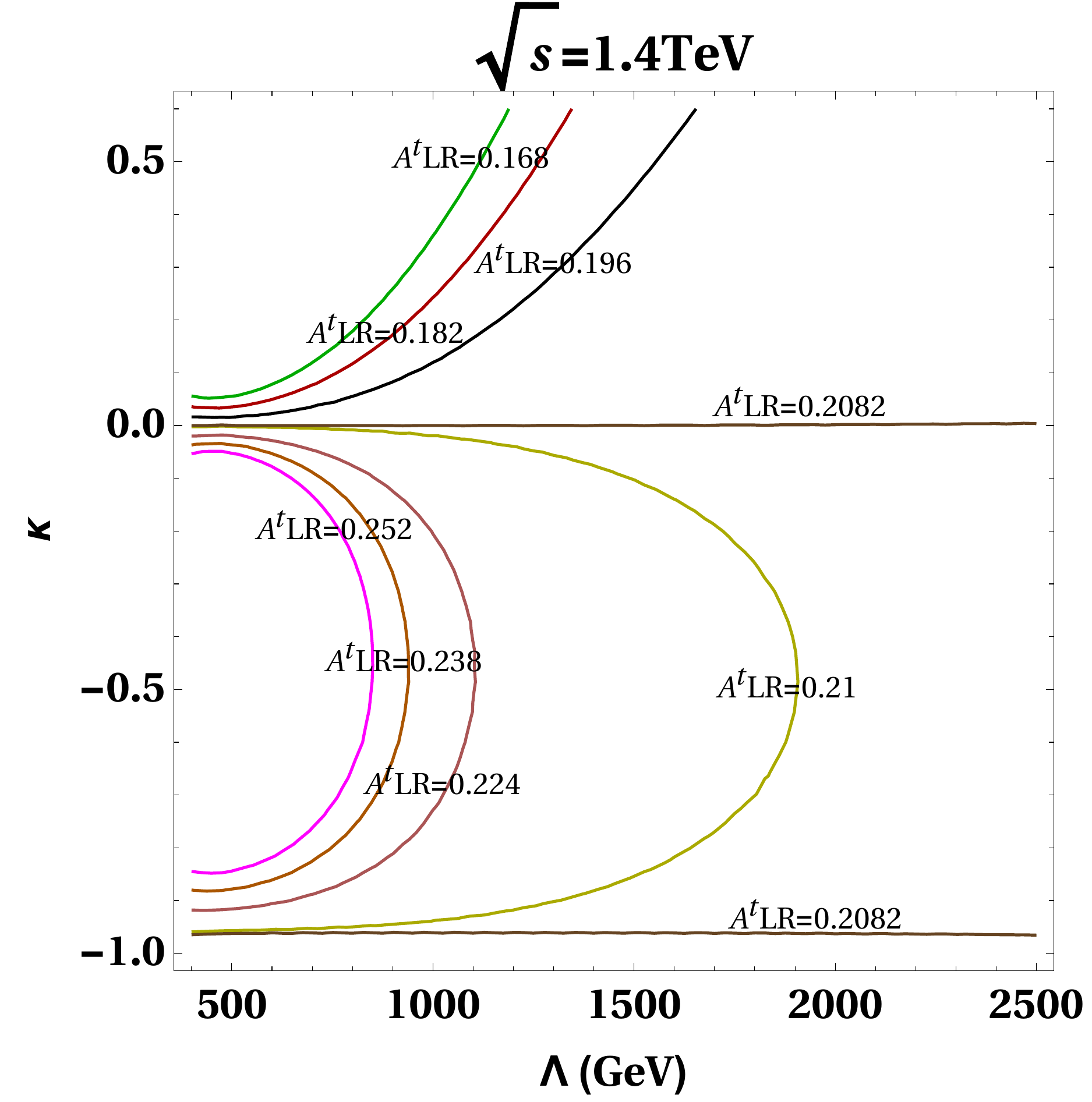}
\hfill
\includegraphics[width=.5\textwidth]{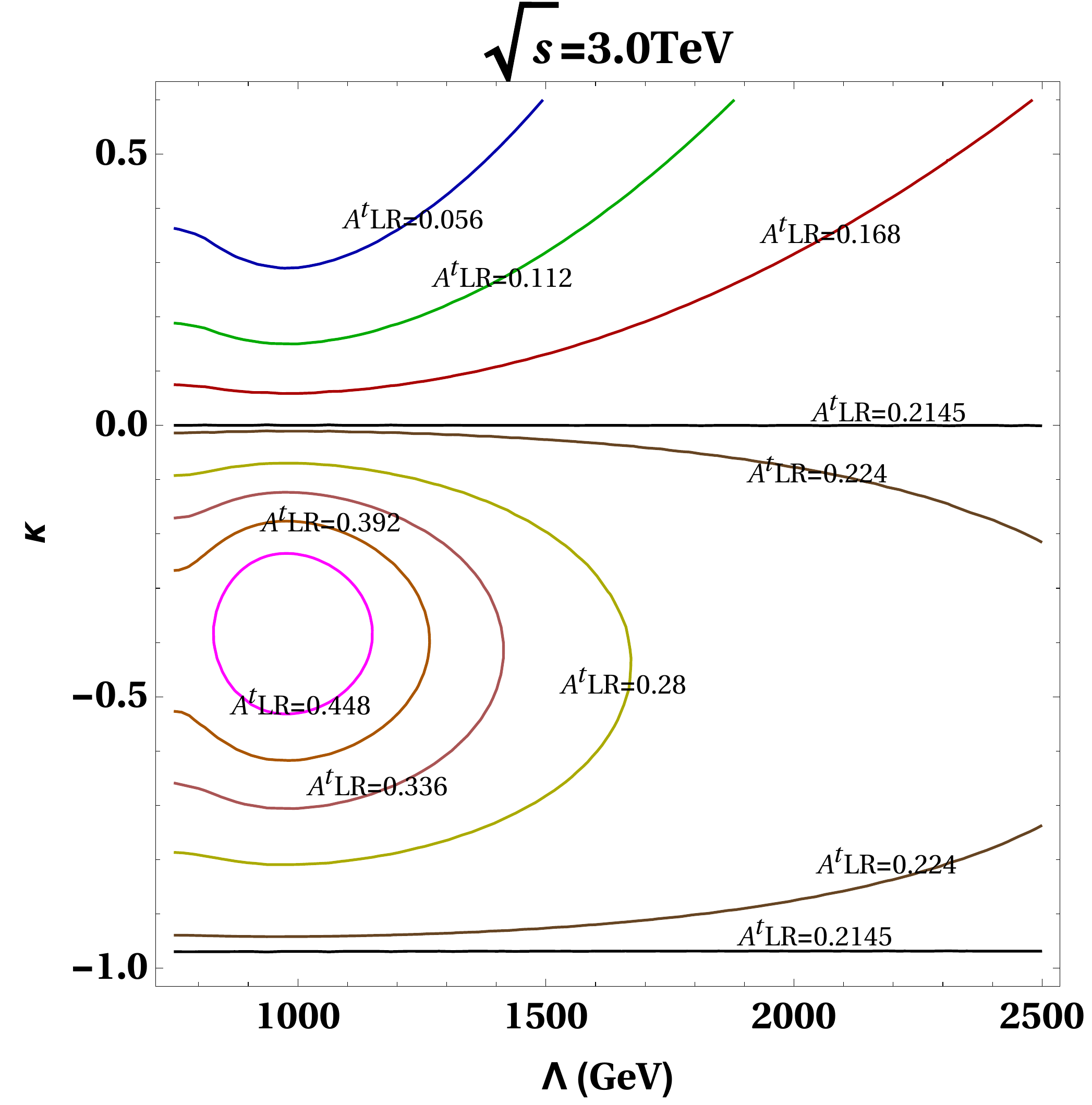}
\caption{\label{fig:11} Final state top quark left-right asymmetry}
\end{figure}
In the non-minimal model, there is no change in the decay rate, however, it gives an insight about non-minimal coupling $\kappa$ and
NC scale $\Lambda$ when left handed top quark produced from $e^{-} e^{+}$ annihilation. Thus it is useful to compute top quark left-right asymmetry\cite{toplrasy1,toplrasy2}.
\begin{equation}
 A^{t}_{LR} = \frac{\sigma (e^{-} e^{+}\longrightarrow t_{L} \overline{t})-\sigma (e^{-} e^{+}\longrightarrow t_{R} \overline{t})}{\sigma (e^{-} e^{+}\longrightarrow t \overline{t})} 
\end{equation}
In the numerator while obtaining cross section, we have summed over the initial state helicity for electron and positron 
as well as final state hecility of the anti-top quark except final state top quark 
helicity, in the denominator, we have summed over all the helicity states on the incoming 
and outgoing particles.
In Fig.\ref{fig:11}, we have shown the contour plots in the $\kappa - \Lambda$ plane corresponding
to different values of left-right asymmetry ($A^{t}_{LR}$) of top quark.
The standard model $A^{t}_{LR}$ increases with the increase in the machine energy, as the production cross-section
for left handed top quark is higher than the right handed top quark production at high 
energy limit. In the SM, one finds the top quark left-right asymmetry 
$20.8\% (21.45\%)$ at the machine energy $\sqrt{s}=1.4 (3.0)$ TeV. In the negative $\kappa$ region, $\kappa_{max}\simeq-0.43$ we get 
the asymmetry $25\%$ (approx) corresponding to $\Lambda=840$ GeV at the machine energy 
$\sqrt{s}=1.4$ TeV. Similarly, for $\sqrt{s}=3.0$ TeV, one find asymmetry upto $28\%$ (approx) 
for $\Lambda=1670$ GeV.
\subsection{Polarized beam analysis}\label{sec:polbeam}
In general, the electron and positron beams can have two types of polarizations which are transverse and longitudinal polarization.
Since we consider the bunch of massless electrons as a beam at linear colliders, the transverse polarizations are negligible,
thus one can define the total cross section with arbitrary longitudinal polarization ($P_{e^{-}},P_{e^{+}}$) given by
\begin{equation}
 \sigma_{P_{e^{-}}P_{e^{+}}} = \frac{(1-P_{e^{-}})}{2}\frac{(1+P_{e^{+}})}{2} \sigma_{LR} + \frac{(1+P_{e^{-}})}{2}\frac{(1-P_{e^{+}})}{2} \sigma_{RL}
\end{equation}
Where $\sigma_{LR}$ is the total cross section of the top pair production when the initial left-handed $e^{-}$ beams and 
initial right-handed $e^{+}$ beams are considered if fully polarized at $P_{e^{-}}=-1 , P_{e^{+}}=+1$. The $\sigma_{RL}$ is defined analogously.
\begin{equation}
 \sigma_{LR}= \sum_{h_{t},h_{\overline{t}}} \sigma_{L\,R\, h_{t}\,h_{\overline{t}}} \, \quad \, \sigma_{RL}= \sum_{h_{t},h_{\overline{t}}} \sigma_{R \,L\, h_{t}\,h_{\overline{t}}}
\end{equation}
Here $h_t$ and $h_{\overline{t}}$ correspond to the helicity states of the top and anti-top quarks, which 
are summed over.The total cross section will enhance when electron and positron beams are polarized and the signs of $P_{e^{-}}$ and $P_{e^{+}}$ are opposite.
This can be realized when the total cross section written in terms of effective polarization as given below\cite{Haber,HelicLC}. 
$$ \sigma_{P_{e^{-}}P_{e^{+}}} = ( 1-P_{e^{-}}P_{e^{+}} ) \sigma_{0} (1-P_{eff} A_{LR})$$
\begin{figure}[h]
\centering 
\includegraphics[width=.5\textwidth]{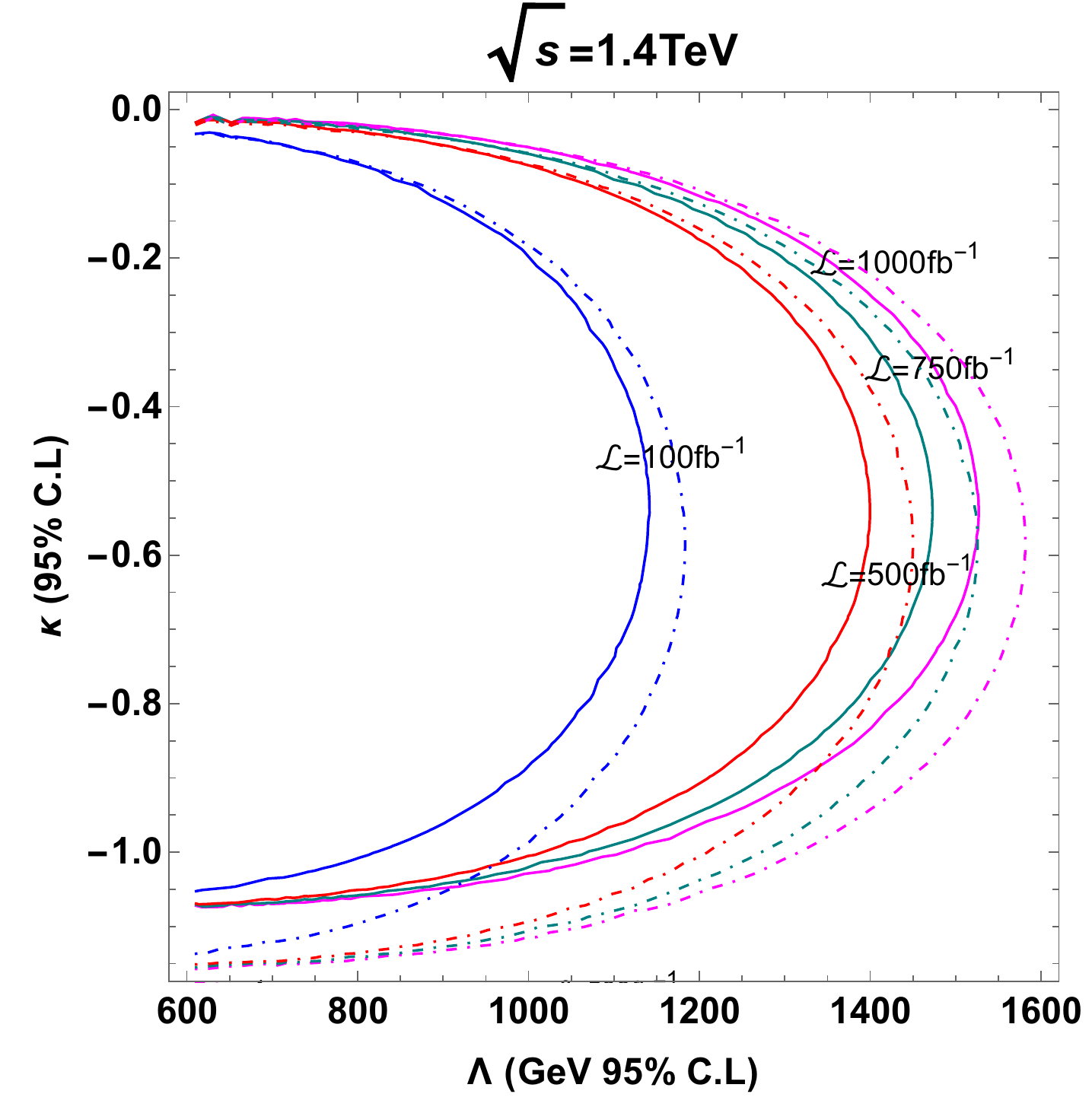}
 \hfill
\includegraphics[width=0.5\textwidth]{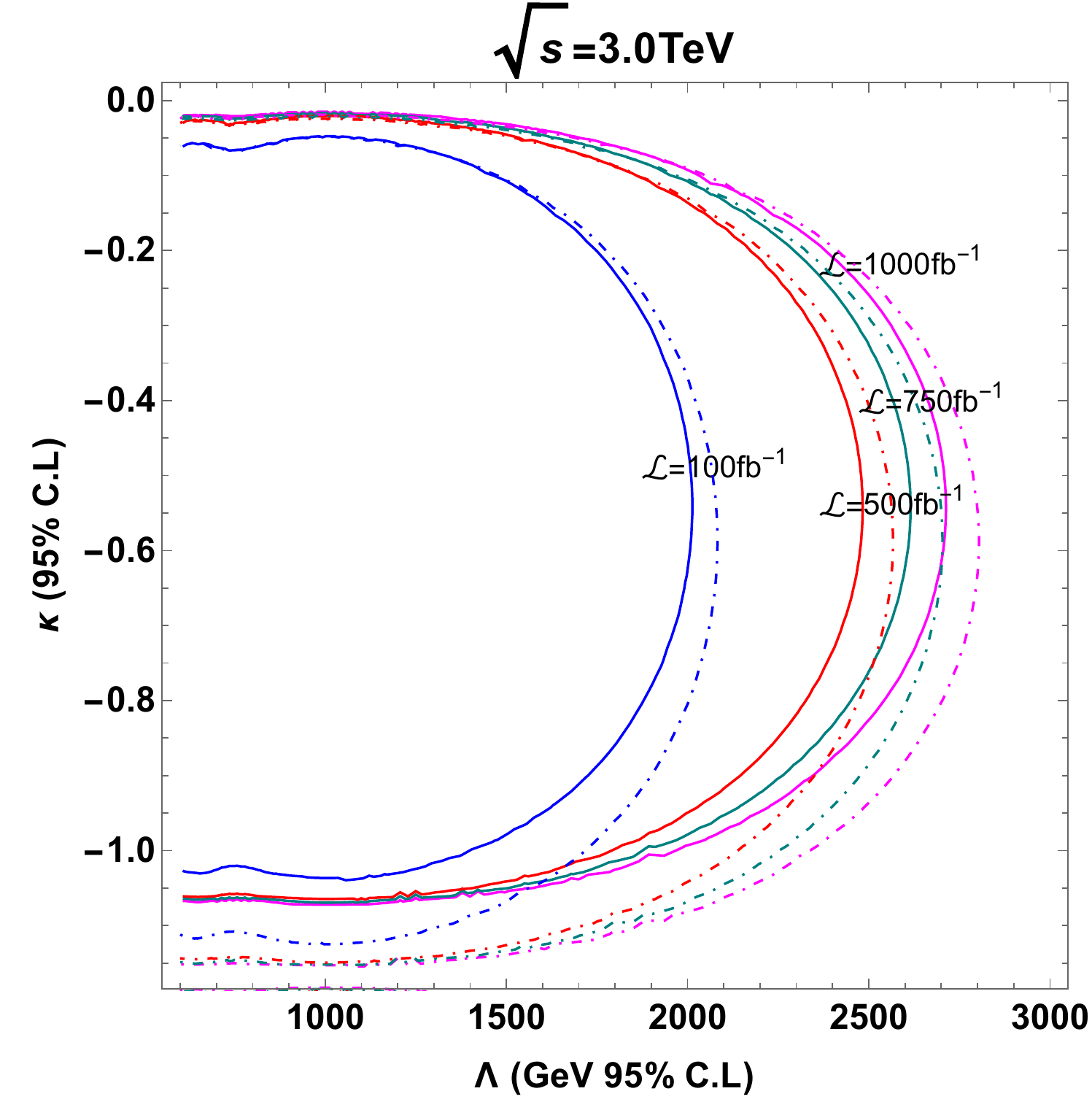}
\caption{\label{fig:12} The figure depicts $\chi^{2}$ statistical test ($ 95\%  C.L$) of polarized the noncommutative signal event
arises due to azimuthal anisotropy only considering the $-^{ve}$ region of the $\kappa$.The color line corresponds to $ P_{e^{-}}=-80\%,P_{e^{+}}=30\%$ and dot-dashed color line corresponds to $ P_{e^{-}}=-80\%,P_{e^{+}}=60\%$ for various 
integrated luminosity $\int \mathcal{L}\,dt =100(\textcolor[rgb]{0,0,1}{blue}),\,500(\textcolor[rgb]{1,0,0}{red}),\,750(\textcolor[rgb]{0,0.5,0.5}{cyan}) \,fb^{-1}$ and $1000(\textcolor[rgb]{1,0,1}{megenta})\, fb^{-1}$ 
at CLIC $\sqrt{s}=1.4$ TeV (left) and $\sqrt{s}=3.0$ TeV (right).}
\end{figure}
\begin{table}[h]
\centering
\resizebox{\textwidth}{!}{%
\begin{tabular}{|l|c|c|c|c|}
\hline	
Integrated luminosity ($\mathcal{L}$) & $100\, fb^{-1}$ & $500\, fb^{-1}$ & $750\, fb^{-1}$ & $1000\, fb^{-1}$\\
\hline 
  \begin{tabular}{l}
$\kappa_{max}=-0.5445$ and $P_{\{e^{-},e^{+}\}}=\{-0.8\,,\,0.3\}$\\
 \hline 
 \hline
 \end{tabular}  & & & & \\
Lower limit on $\Lambda$: $\sqrt{s}=1.4$TeV  & $1.131\, TeV$  & $1.390\, TeV$ & $1.462\, TeV$ & $1.517\, TeV$ \\ 
& & & & \\
Lower limit on $\Lambda$: $\sqrt{s}=3.0$TeV  &  $2.000\, TeV$ & $2.474\, TeV$  & $2.606\, TeV$ & $2.703\, TeV$ \\ 
\hline
 \begin{tabular}{l}
 $\kappa_{max}=-0.607$ and $P_{\{e^{-},e^{+}\}}=\{-0.8\,,\,0.6\}$\\
 \hline 
 \hline
 \end{tabular}
 & & & & \\
Lower limit on $\Lambda$: $\sqrt{s}=1.4$TeV  & $1.172\, TeV$  & $1.440\, TeV$ & $1.514\, TeV$ & $1.570\, TeV$ \\ 
& & & & \\
Lower limit on $\Lambda$: $\sqrt{s}=3.0$TeV  &  $2.075\, TeV$ & $2.560\, TeV$  & $2.70\, TeV$ & $2.80\, TeV$ \\ 
\hline 
\end{tabular}}
\caption{\label{tab:2} The lower bound on noncommutative scale $\Lambda$ ($95\%$C.L) and $\kappa_{max}=\{-0.5445\,,\,-0.607\}$ obtained by $\chi^{2}$ analysis 
when  $P_{\{e^{-},e^{+}\}}=\{-0.8\,,\,0.3\}$ and  $P_{\{e^{-},e^{+}\}}=\{-0.8\,,\,0.6\}$ for four different integrated luminosity.}
\end{table}
Here the unpolarized total cross section $ \sigma_{0}= (\sigma_{LR}+\sigma_{RL})/4$, left-right asymmetry $ A_{LR}=(\sigma_{LR}-\sigma_{RL})/(\sigma_{LR}+\sigma_{RL})$
and effective polarization $P_{eff}= (P_{e^{-}}-P_{e^{+}})/(1-P_{e^{-}}P_{e^{+}})$. The $P_{eff}$ can be achieved $-1$ when
$P_{e^{-}}=-1.0\,,P_{e^{+}}=0.0$ and $P_{e^{-}}=-1.0\,,P_{e^{+}}=1.0$. So that the total cross section enhanced by
$\sigma_{\{-1,0\}}=\sigma_{0}+\sigma_{0} \,A_{LR}$ and $\sigma_{\{-1,1\}}=2(\sigma_{0}+\sigma_{0} \,A_{LR})=2\sigma_{\{-1,0\}}$ respectively. 
In the future linear colliders, the polarization can be studied with $P_{e^{-}}=-0.8\,,P_{e^{+}}=0.3$ and $P_{e^{-}}=-0.8\,,P_{e^{+}}=0.6$ which will 
enhance the total cross section significantly by following expressions 
\begin{eqnarray}
 \sigma_{\{-0.8,0.3\}} & = & 1.24\,\sigma_{0}+1.1\,\sigma_{0} \,A_{LR}\\
 \sigma_{\{-0.8,0.6\}} & = & 1.48\,\sigma_{0}+1.4\,\sigma_{0} \,A_{LR}
\end{eqnarray}
Note that the above set of equations are same for all new physics as well as for the standard model. 
The beam polarization enhances the new physics signal ($S$) and suppresses the background ($B$) rates by 
significant increment of $S/B$ or $S/\sqrt{B}$.
\par In our analysis $\sigma_{0}$ and $A_{LR}$ are function of $\Lambda$. We made $\chi^{2}$ test for polarized beam 
analysis by keeping the azimuthal anisotropy as an observable. The polarization enhances the lower limit on
NC scale and non minimal coupling $\kappa_{max}$ also. The lower bound on noncommutative scale is given 
in the table.\ref{tab:2} for four values of integrated luminosity.
\begin{figure}[h]
\centering 
\includegraphics[width=0.55\textwidth]{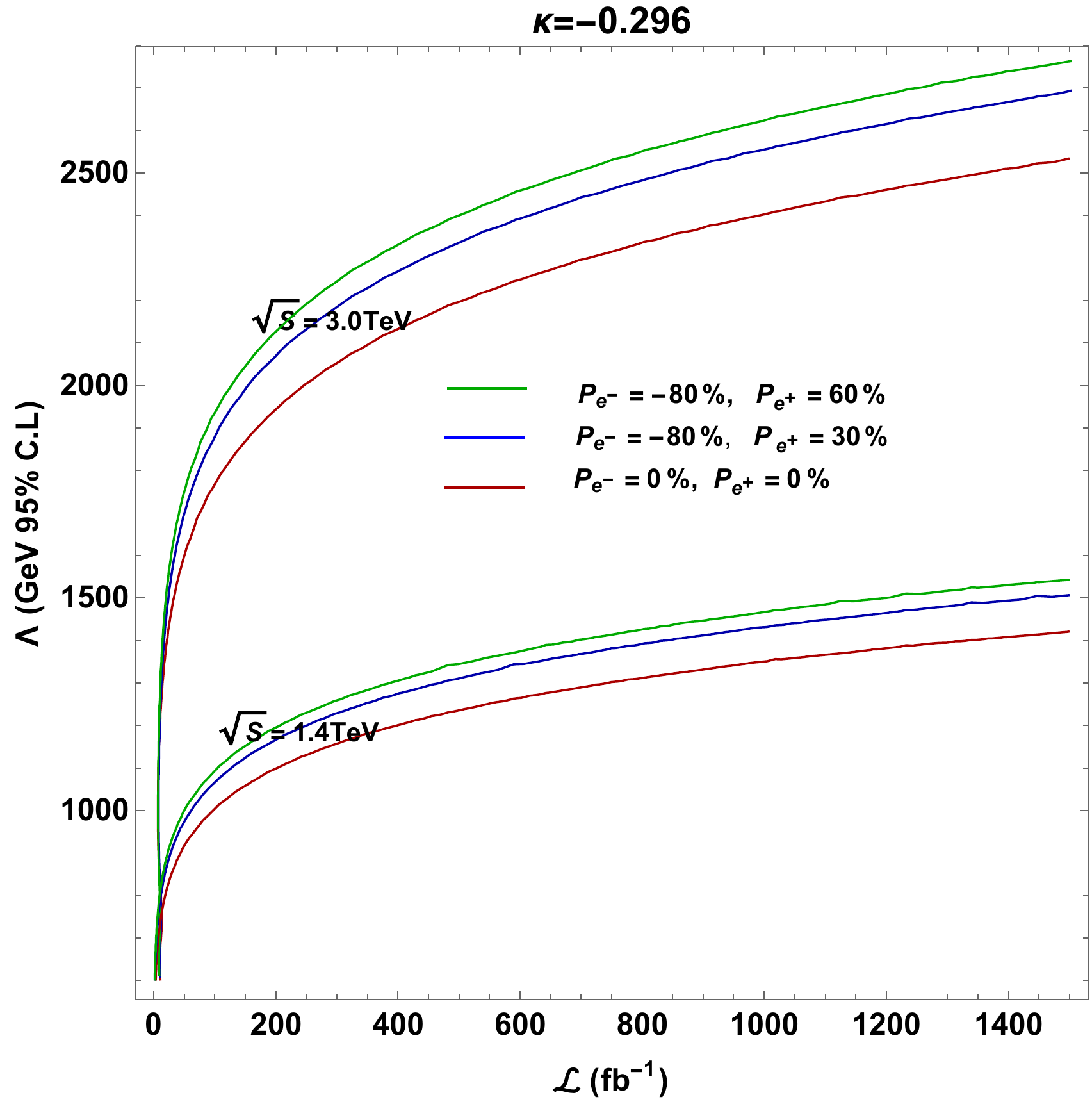}
\caption{\label{fig:13} The figure depicts $\chi^{2}$ statistical test ($ 95\%  C.L$) of polarized noncommutative signal event
arises due to azimuthal anisotropy.}
\end{figure}
\par The sensitivity of the NC scale $\Lambda$ on integrated luminosity of the polarized electron positron beam has given in the Fig.\ref{fig:13}.
In the above Fig.\ref{fig:12} we have seen that the polarization has pushed $\kappa_{max}=-0.296$ into $\kappa_{max}=-0.607$ when $P_{\{e^{-}\,,\,e^{+}\}}=\{-0.8\,,\,0.6\}$.
But one can compare the lower bound on NC scale with the unpolarized beam  and partially polarized beam luminosity for fixed value of $\kappa_{max}$ at $-0.296$ which is shown in the 
Fig.\ref{fig:13}. There are significant enhancement in the lower limit of the NC scale. If any one of the beam fully polarized and another one has unpolarized (fully polarized) 
which gives $P_{eff}=100\%$ and one can get two (four) times enhancement in the total cross section.

\section{Summary and conclusion}\label{sec:conclude}
We study the top quark pair production $ e^+ e^- \stackrel{\gamma,Z}{\longrightarrow} t \overline{t} $ at the TeV energy 
 linear collider in the non-minimal NCSM within the framework of covariant $\Theta$-exact Seiberg-Witten approach. 
 Although the NC tensor $\Theta_{\mu\nu}$ breaks the rotational invariance due to $\Theta_{0i}\neq 0$, however, 
 it preserves the translational symmetry which is invariant under the $T(2)$ subgroup of VSR. We study the NC cross section correction
 observable $\Delta \sigma = \sigma^{NC} - \sigma^{SM}$ corresponding to the machine energy 
 $\sqrt{s} = 1.4$~TeV as well as $\sqrt{s} = 3.0$~TeV, and presented our result in terms of the non-minimal coupling 
 $\kappa$ and the NC scale $\Lambda$ corresponding to positive and 
 negative $\Delta \sigma$. For $\kappa_{max} = -0.296$, we find the lower bound on the NC scale 
 $\Lambda \ge 780(1680)$ GeV corresponding to $\Delta \sigma \sim 10\%$ at the machine energy 
 $\sqrt{s} = 1.4(3.0)$ TeV. For a specific value of $\Lambda$ at a fixed positive coupling $\kappa$, we found a 
 specific optimal collision energy relation which is $\sqrt{s_0} = 2.52105~\Lambda + 39.443$, may be quite useful to look for the signature of the spacetime noncommutativity. The normalized 
 azimuthal distribution is found to vary according to $cos^2[f(1/\Lambda^2) sin\theta cos\phi]$  for 
 $0 > \kappa > -0.596$ and $sin^2[f(1/\Lambda^2) sin\theta cos\phi]$  for $\kappa > 0$ and 
 $\kappa < -0.596$. A statistical $\chi^2$ analysis of the azimuthal anisotropy gives rise a lower bound on 
 $\Lambda = 1.0,~1.2,~1.3,~1.35$ TeV corresponding to linear collider luminosity ${\mathcal{L}} = 100,~500,~750,~1000~\rm{fb^{-1}}$ 
 for $\kappa_{max} = -0.296$ at the machine energy $\sqrt{s} = 1.4$ TeV. \\
  \noindent We perform a detailed helicity analysis of the $t~\overline{t}$ production. We present the helicity 
  correlation $C_{t{\overline{t}}}$ of the final state top quark and  anti-top quark produced with a certain 
  helicity and found that such a correlation which is constant at $ \sqrt{s}=1.4(3.0)$ TeV i.e. $C^{SM}_{t{\overline{t}}} = -0.9518\, ,\,(-0.9894) $ respectively, varies with the 
  NC scale $\Lambda$ for different coupling constant $\kappa$ in the NCSM.We investigate the sensitivities of the 
  top quark left-right asymmetry $A^t_{LR}$ on the non-minimal coupling $\kappa$ and the NC scale $\Lambda$ 
  and find that the asymmetry is about $25\%$ in the non-minimal NCSM with $\kappa_{max} = -0.43$ 
  corresponding to $\Lambda = 840~$ GeV at the machine energy $\sqrt{s} = 1.4$ TeV.  Further, we perform a 
  detailed $\chi^2$  analysis for the polarized electron-polarized beam with $P_{e^-} = -80\%$ and $P_{e^+} = 60\%$ corresponding to 
  the machine energy $\sqrt{s} = 1.4~(3.0)$ TeV for different machine luminosities. We obtain the lower bound on 
  $\Lambda = 1.17(2.08),~1.44(2.56),~1.51(2.70),~1.57(2.80)$ corresponding to the machine luminosity 
  ${\mathcal{L}} = 100,~500,~750,~1000~\rm{fb^{-1}}$ for $\kappa_{max} = -0.607$. Finally, we studied the intriguing mixing of the UV and the IR by invoking a specific structure of noncommutative anti-symmetric tensor $\Theta_{\mu\nu}$ 
 which is invariant under translational $T(2)$ VSR Lorentz sub group given in the appendix \ref{app:uvir}.


\appendix
\section{Removal of UV/IR mixing and NC helicity amplitude calculation}
\subsection{Removal of UV/IR mixing in the one loop calculation}\label{app:uvir}
There are several attempts has been made to solve/remove the UV/IR mixing which are appears as phase factor due to Moyal product between NC fields
in the spacetime noncommutative field theory. In the $\Theta-$expanded Seiberg-Witten map approach, the UV/IR mixing disappears \cite{MWURIR1,SWURIR5} and the renormalibility
has under control up to one loop $\mathcal{O}(\Theta)$. But in the case of $\Theta-$exact Seiberg-Witten map approach, the UV/IR mixing are present until the noncommutative tensor $\Theta^{\mu\nu}$
takes the special structure. The specific structure of the $\Theta^{\mu\nu}$ removes the both singularity in the $\Theta-$exact as well as covariant $\Theta-$exact 
noncommutative field theory which is given in \cite{SWURIR8,HKT3,SWURIR9}. But those structure of $\Theta^{\mu \nu}$ given in 
\cite{SWURIR8,SWURIR9} violates the VSR $T(2)$ translation property.
\par Here, we have shown that our choice of $\Theta^{\mu \nu}$ removes the UV/IR mixing which satisfy the translational invariant under
VSR group. Since $\Theta^{0i}\neq 0$ in our case, thus the azimuthal anisotropy play an unique feature to probe the spacetime noncommutativity at particle colliders.\\
In order to remove the UV/IR mixing at one-loop self energy correction, R. Horvat et.al.,\cite{SWURIR9,HKT3} introduced non-minimal fermion coupling ($\kappa$) in the $U(1)$ gauge theory following manner
\begin{equation}
 S^{U(1)}_{f}= \int -\frac{1}{4} f_{\mu\nu}f^{\mu\nu}+i \overline{\psi} \slashed{\partial} \psi - i \Theta^{ij} \overline{\psi} \gamma^{\mu}\left( \frac{1}{2}f_{ij}\circledast \partial_{\mu}\psi-\kappa f_{\mu i}\circledast \partial_{j}\psi \right)
\end{equation}
Where $ f_{\mu\nu}=\partial_{\mu}a_{\nu}-\partial_{\nu}a_{\mu}$. Eventually, we get following feyman rule for massless neutral fermion-fermion- photon interaction
\begin{eqnarray}
-e\widetilde{F}_{\circledast}(p,q) 
 \left[\kappa\left( \slashed{q}(\Theta p)^{\mu} -(q\Theta p) \gamma^{\mu}\right) -(\Theta q)^{\mu} \slashed{p} \right] \label{feynrulec}
\end{eqnarray}
Here $p$ is the fermion momentum which flows towards vertex and $q$ is the photon incoming momentum. Note that 
$$[f,g]_{\star}=i \Theta^{ij} \partial_{i}f \circledast \partial_{j}g$$
In principle,
$$ [A_{\mu},\psi]_{\star} = A_{\mu}\star \psi - \psi \star A_{\mu} $$
Thus the covariant derivative for neutral fermion can be written as 
$$ D_{\mu} \psi = \partial_{\mu} \psi - i \kappa\,[A_{\mu},\psi]_{\star}$$
When $\kappa=0$, we get the commutative interaction which is not consider in eqn. \ref{feynrulec}. We can realize that $\kappa$ is the non-minimal NC coupling, the strength of the neutral fermion and 
abelian gauge field interaction is proportional to $\kappa\, e$, which is shown in eqn.\ref{photonfeyn}.
We follow the $SU_{\star}(2)_{L} \otimes U_{\star}(1)_{Y}$ fermion action which is given in \cite{HKT1,HKT3}, allows the interaction between
neutral right handed fermion and abelian gauge field in the constant anti-symmetric background field.
\begin{figure}[h]
\centering 
\includegraphics[width=0.55\textwidth]{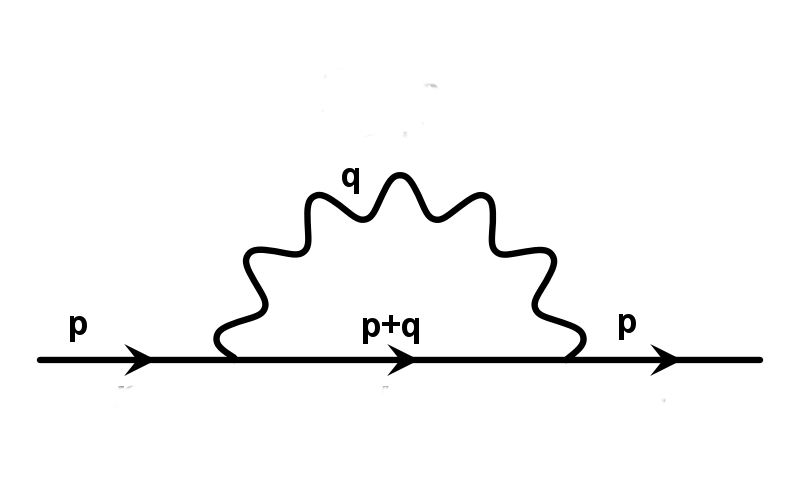}
\caption{\label{fig:14} Neutral fermion (Neutrino) one-loop self energy Feynman diagram ($\Sigma_{1}$).}
\end{figure}
Let us consider the massless neutral fermion one-loop self energy correction in the covariant $\Theta-$exact NCSM as follows
\begin{eqnarray}
 \Sigma_{1} & = & -(\kappa\,e)^{2}\mu^{4-D} \int \frac{d^{D}q}{(2\pi)^{D}} \, \frac{-i \eta_{\mu \nu}}{q^{2}}\,\widetilde{F}_{\circledast}(p,q)\,
 \left[ (q\Theta p) \gamma^{\mu} + \slashed{p} (\Theta q)^{\mu} -\slashed{q}(\Theta p)^{\mu}\right] \, \frac{i \slashed{p}+\slashed{q}}{(p+q)^{2}} \nonumber \\
 & & \widetilde{F}_{\circledast}(p,q) \left[ -(q\Theta p) \gamma^{\nu} - \slashed{p} (\Theta q)^{\nu} +\slashed{q}(\Theta p)^{\nu}\right] \label{self}
\end{eqnarray}
Where $ \widetilde{F}_{\circledast}(p,q) = 2 \left( \frac{\sin((q\Theta p)/2)}{q\Theta p} \right) $. Our results and ref \cite{HKT3} results are differs only by multiplication factor which is $\kappa^{2}$. 
Apart from bubble diagram other tadpole one-loop diagrams ($\Sigma_{3}\, ,\,\Sigma_{4}$) vanishes due to charge conjugation symmetry ($\Theta^{\mu\nu}_{C}= -\Theta^{\mu\nu}$)
as shown in ref \cite{HKT3}. The four field tadpole (2-fermion 2-photon) one-loop ($\Sigma_{2}$) vanishes which can be easily viewed when $ \eta_{\mu\nu} V^{\mu\nu}_{3}(p,-p,q,q)=0 $, 
the four field interaction $V^{\mu\nu}_{3}(p,-p,q,q)$ can get from  eqn (2.10) in ref{\cite{HKT3}}. Eventually, the non-vanishing bubble diagram (as shown in fig.\ref{fig:14}) has to be evaluate.
We follow
\begin{itemize}
 \item presence of $q\Theta p$ dimensionless scalar quantity in the denominator, we use HQET parametrization \cite{HQET} to combine all quadratic and linear denominator.
 \item we use Schwinger technique to turn the denominator into Gaussian integrals and absorb NC phase factor into that.
 \item we calculate all the integrals which are presented in the numerator by using \cite{Gradshteyn}.
\end{itemize}
In field theory, if the Hamiltonian has positivity properties then we can analytically continue to imaginary values of time along adopting the Weyl's unitarity trick
which leads Euclidean field theory of causal propagation with positive energies. The analytic continuation is defined by 
the rotation of the vector in the complex plane an angle of $\pi/2$ about the origin namely Wick rotation. Therefore one can go from 
Minkowski spacetime to Euclidean spacetime by replacing the time $t$ by $i\,t$ and the  energy $p^{0}$ by $i\,p^{0}$.
In our NCQFT, the analytic continuation demands that the noncommutative space-time (NCST) component $\Theta^{0i}$ undergo $\Theta^{0i} \rightarrow -i \Theta^{0i}$ and leaves 
invariant Moyal phase ($q \Theta p$) under Wick rotation \cite{JT1,JT2}. Here the time component of four momenta undergo $q^{0} \rightarrow i q^{0}$ and $p^{0} \rightarrow i p^{0}$ by Wick rotation, $q$ is 
loop momenta and $p$ is external momenta. The noncommutative space-space (NCSS) component remains same like three-momenta ($p^{i}$) by Wick rotation. After Schwinger 
parametrization, we get the same Euclidean integrals which is shown in ref \cite{HKT3}. 
The integrals $I_{123}$ equation (A.1), $I_{456}$ equation (A.7) and $I_{789}$ equation (A.8) in ref \cite{HKT3} has to satisfy the important unitarity condition
thereby the Euclidean term $(\widetilde{p})^{2}(=(\Theta p)^{2})$ should be positive. Such term arises from the Schwinger parametrization and 
redefined loop momenta $l$. The loop momenta $l$ is still invariant by Wick rotation using the dual property of VSR NC tensor 
i.e duality of NCST ($\Theta^{0i}$) and NCSS ($\Theta^{3i}$). We can back to Euclidean to Minkowski by analytic continuation. 
In order to do that, we have to prove $(\widetilde{p})^{2}$ is positive in Minkowski spacetime,
otherwise the integral would not converge and can't achieve unitarity \cite{JT1,JT2}. Our case we get,
     $$\widetilde{p}^{2} = \frac{1}{\Lambda^{4}}(a^{2}+b^{2}).(p_{0}-p_{3})^{2} = \theta^{2}p_{-}^{2}\, \, \Longrightarrow \, \, \text{Always positive} $$
Here $a$ and $b$ are real and $\theta^{2}=\frac{1}{\Lambda^{4}}(a^{2}+b^{2})$ and $p_{-}$ is the back-ward lightcone momentum of the propagating neutral particle. Since $\widetilde{p}^{2}$ 
is positive, the Light like spacetime VSR noncommutativity obeys the unitarity property perturbatively.
After long calculation we get simplified equation \ref{self} for $ D = 4- \epsilon$ and in $\epsilon \rightarrow 0$ limit,
\begin{equation}
 \Sigma_{1} = -(\kappa\,e)^{2}\gamma_{\mu}\left[ p^{\mu} \mathcal{A} + (\Theta \Theta p)^{\mu} \frac{p^{2}}{(\Theta p)^{2}} \mathcal{B} \right]
\end{equation}
Here
\begin{eqnarray}
\mathcal{A} & = & \frac{1}{(4\pi)^{2}} \left[ (S_{1}+2S_{2})\mathcal{A}_{1} + (1+S_{1}+S_{2})\mathcal{A}_{2} \right] \\
\mathcal{B} & = & -\frac{2}{(4\pi)^{2}} \mathcal{A}_{2}
\end{eqnarray}
\begin{eqnarray}
 \mathcal{A}_{1} & = & \frac{2}{\epsilon} + \ln\left( \pi e^{\gamma_{E}}\mu^{2}(\Theta p)^{2} \right)
  +\frac{1}{2} \sum_{k=1}^{\infty} \frac{\sqrt{\pi}}{\Gamma \left( k+1 \right) \Gamma \left(k+\frac{3}{2} \right) } \left( \frac{p^{2}(\Theta p)^{2}}{16}\right)^{k} \mathcal{A}_{3} \nonumber \\
 \mathcal{A}_{2} & = & 2-\frac{1}{2} \sum_{k=0}^{\infty} \frac{\sqrt{\pi} \, \Gamma \left(k+\frac{1}{2} \right)}{\Gamma \left( k+1 \right) \Gamma \left(k+\frac{3}{2} \right) \Gamma \left(k+\frac{5}{2} \right)} 
 \left( \frac{p^{2}(\Theta p)^{2}}{16}\right)^{k+1} \left( \mathcal{A}_{3}+ \psi_{0}\left( k+\frac{1}{2} \right) - \psi_{0}\left(k+\frac{5}{2} \right) \right) \nonumber \\
 \mathcal{A}_{3} & = & \ln\left( \frac{p^{2}(\Theta p)^{2}}{16} \right)- \psi_{0}\left( k+1 \right) - \psi_{0}\left(k+\frac{3}{2} \right)
  \end{eqnarray}
Where $\psi_{0}(z)$ are di-gamma functions and $\gamma_{E} \simeq -0.577215$ is Euler-Mascheroni constant.
Further $S_{1}$ and $S_{2}$ are scale independent $\Theta$ ratios. We get the scalar quantity $\Theta_{\mu \nu} \Theta^{\mu \nu}$ in $S_{1}$ which arises after integrating the loop momenta.
\begin{equation}
 S_{1} = p^{2} \frac{\Theta_{\mu \nu} \Theta^{\mu \nu}}{(\Theta p)^{2}} \,; \quad \, S_{2} = p^{2} \frac{(\Theta\Theta p)^{2}}{(\Theta p)^{4}};
\end{equation}
We know that $(\Theta p)^{2}$ is always positive, then the above equations are valid in the Minkowski spacetime, Which means that
the Euclidean 4D space results are same in the Minkowski 4D spacetime by analytic continuation according to Weyl's trick in field theory\cite{JT1,JT2}. 
By looking at the equation \ref{lightNC}, one can conclude that $S_{1}=0$ and by using equation \ref{nctensor}, we get $(\Theta\Theta p)^{2} = 0$ for all value of $p$.
The UV/IR mixing and hard $1/\epsilon$ UV divergence vanishes but IR divergence still present in the NC theory.
\begin{equation}
 \Sigma_{1} = (\kappa\,e)^{2}\frac{1}{(4\pi)^{2}}\left[ \slashed{p} -2 \slashed{\doubletilde{p}} \frac{p^{2}}{(\Theta p)^{2}} \right]\mathcal{A}_{2}
\end{equation}
Here $ \doubletilde{p}^{\mu} = (\Theta \Theta p)^{\mu} = \Theta^{\mu\nu} \Theta_{\nu\rho} p^{\rho} $. Interestingly, $\slashed{\doubletilde{p}}$ can be written as $\frac{(\doubletilde{p})^{2}}{\slashed{\doubletilde{p}}}$, 
and $(\doubletilde{p})^{2}=0 \, \forall \, p$ by VSR, thereby the quadratic IR singularity disappears. At last, we have arrived the logarithmic IR singularity with convergent sum of $f^{k}(p,\widetilde{p})$ as follows,
\begin{equation}
 \Sigma_{1} = (\kappa\,e)^{2}\frac{\slashed{p}}{8\pi^{2}}\left[ 1 - \frac{\sqrt{\pi}}{4} \, \left( \frac{p^{2} \widetilde{p}^{2}}{16} \right) \left\{ \ln\left( \frac{p^{2} \widetilde{p}^{2}}{16} \right) 
 \sum_{k=0}^{\infty} \, f^{k}(p,\widetilde{p}) + \sum_{k=0}^{\infty} \, \psi_{0}^{k} \, f^{k}(p,\widetilde{p}) \right\}\right] 
\end{equation}
$$ \text{Here} \, f^{k}(p,\widetilde{p}) =  \frac{\left( \frac{p^{2} \widetilde{p}^{2}}{16} \right)^{k}\, \Gamma \left(k+\frac{1}{2} \right)}{\Gamma \left( k+1 \right) \Gamma \left(k+\frac{3}{2} \right) \Gamma \left(k+\frac{5}{2} \right)} 
\, ,\, \, \text{ and} \, \, \psi_{0}^{k} \, \, \text{is related to di-gamma function, } $$
$$ \text{which is } \, \, \psi_{0}^{k} = \psi_{0}( k+1/2 ) - \psi_{0}( k+1 ) - \psi_{0}(k+3/2)- \psi_{0}(k+5/2 )$$
The IR divergence pay the attention on motion of the particle inside
noncommutative plane which results that the spatial extension $\Theta p$ cannot shrink to zero \cite{SWURIR8}. On the other hand, the IR singularity can be removed by
re-summing the propagator analogous to finite temperature field theory and introducing appropriate counter term in the interaction Lagrangian. We will discuss more in the near future work.
\par In the scalar field theory, the mass of the scalar is far from UV scale, thus one can think of the naturalness of the scalar mass UV physics 
knows nothing about the theory in the far IR. Instead of introducing the new symmetries and extra-dimensions one might hope that connecting the 
far IR and far UV can solve the naturalness problem \cite{craig}. Interestingly, the NCQFT and quantum gravity exhibit the UV/IR mixing by its nature. Recently proposed 
weak gravity conjecture (WGC) \cite{vafa,WGC} on the form of hierarchical UV/IR mixing restricts the mass of the scalar to a IR scale far below 
the UV scale which is associated to quantum gravity. Here we have given a glimpse of WGC on non-vanishing NC scale parameter $\Theta$ which was considered in \cite{qhuang,mli}. 
Consider the construction of the soliton in $2+1$ dimensional NC scalar field theory, 
in general $2l+1$ dimensional theory\cite{soliton1,soliton2} one can understand the feature of the NC parameter $\Theta$. Where $l$ is $1,2,...$
According to Derrick theorem, the kinetic and potential energies are decreases when all 
length scale are shrinks as $L\rightarrow \lambda L$, consequently no finite-size minimum can occur at $\lambda\rightarrow 0$. But in the NC scenario, the Derrick theorem fails
by existence of classically stable GMS soliton \cite{soliton1,soliton2} in the presence of distinguished length scale $\sqrt{\Theta}$ especially in $2+1$ spacetime. 
Here $\Theta$ is the eigen value of the $\Theta^{\mu \, \nu}$. In the weak gravity limit, the soliton is presented in the two-dimensional consistent quantum gravity system
which is of reduction in the higher dimension. If the gravity coupling increases these soliton disappears then there are no more massive particle and S-matrix breaks down.
Thereby the quantum theory survives only in the weak gravity region. Based on Nima Arkani-Hamed et.al., weak gravity conjecture\cite{vafa}, the authors in ref \cite{qhuang,mli}
proposed the NC version of weak gravity conjecture for scalar field theories which is \textit{the effect of gravity would give a much smaller correction to the scalar field theory}, 
thus in higher dimension
\begin{equation}
 \left( \frac{\lambda}{\mu^{2}}\right)^{\alpha} \gtrsim G \label{NCWGC}
\end{equation}
Thereby we get the solution for $\Theta$ when the massive soliton will generate 
a deficit angle in the metric which satisfies $ (8\pi\, G \, E) < 2\pi$ is
$$ \frac{1}{G} \gtrsim \frac{\Theta \, \mu^{2}}{(\lambda /\mu^{2})^{\alpha}} \geqslant \left( \frac{\mu^{2}}{\lambda}\right)^{\alpha}.$$
Where $ E $ is the soliton energy, $G$ is the Newton's constant which is proportional to $M_{\text{pl}}^{-2}$ in $D$ dimension and $\mu \, , \, \lambda$ and 
$\alpha = 2/(n-2)$ are mass of the scalar, coupling of $n$ interacting scalar and $\mathcal{O}(1)$ parameter for $n=4$ in $D$ dimension respectively.
It is naturally gives SM electroweak scale is lower than Planck scale i.e. $(\mu/\sqrt{\lambda}) \lesssim M_{\text{pl}}$ assuming in the $3+1$ spacetime and when $\Theta$ takes
$$ \Theta \geqslant \frac{1}{\mu^{2}}$$
Though the results are trivial but there is no supportive evident for this conjecture \cite{mli} when the NC soliton to be presented in the $3+1$ dimensional 
spacetime. But the conjecture \ref{NCWGC} is similar to WGC given in ref \cite{vafa} for scalar field theories. Although the $2+1$ and $3+1$ dimensional spacetime solitonic solution turns into equivalent when considering the $\phi^{3}$ scalar interaction which is shown in \cite{mli} 
i.e $ \Theta \geqslant \frac{1}{\mu^{2}}$. In addition, the local degrees of freedom per spacetime point for $D$ dimensional spacetime is $D(D-3)/2$. Thus, there is 
no propagation of the gravity(massless) in $D=3$ dimensional spacetime. 
\subsection{NC differential cross section}\label{app:ncxsec}
Here we present the covariant $\Theta$-exact NCSM differential calculation by using VSR T(2) invariant antisymmetric tensor $\Theta_{\mu \nu}$.
One can calculate the differential cross section by using the Feynman rule given in (\ref{photon1},\ref{zboson1},\ref{photon2},\ref{zboson2}). \\
For the photon mediated process, one finds 
\begin{equation}
 \frac{d\sigma^{\gamma}}{d\Omega} = \left(\frac{\alpha^2 \beta A_{f} Q_{e}^{2}}{16 s} \right) \left((1+\cos^{2}\theta)+\left(\frac{4 m^2}{s}\right)\sin^{2}\theta \right)
 \left(Q_{t}^{2}+ f_{\gamma}(\kappa,\Lambda,\sqrt{s})\right) \label{diffphoton}
\end{equation}
where the NC contribution stems from 
 \begin{equation} 
 f_{\gamma}(\kappa,\Lambda,\sqrt{s})=4\kappa(\kappa+Q_{t}) \sin^{2}\left(\frac{p_{4}\Theta p_{3}}{2}\right) 
 \end{equation}
 The signature of the noncommutativity arises from the term  
 $\frac{p_{4}\Theta p_{3}}{2} = \left( \frac{s\, \beta}{4 \Lambda^{2}} \right) \sin\theta \, \cos\phi$. 
 Here $\theta$ is the polar angle which is defined w.r.t the electron beam axis i.e along the $z$-axis
 and $\phi$ is the azimuthal angle. The factor $ \beta=\sqrt{1-\frac{4 m_t^{2}}{s}}$ and $s$ is the squared 
 c.o.m energy. \\
\noindent Similarly, the differential cross section for the $Z$ boson mediated $s$ channel 
process can be written as 
 \begin{eqnarray}
   \frac{d\sigma^{Z}}{d\Omega} & = & \left(\frac{\alpha^2 \beta A_{f} }{16 \sin^{4}2\theta_{w}((s-m_{Z})^{2}+(\Gamma_{Z} m_{z})^{2})} \right)
 \Biggl\{ f_{SM}^{Z}+f_{Z}(\kappa,\Lambda,\sqrt{s}) \Biggr\}
 \end{eqnarray}
where
  \begin{eqnarray}
  f_{SM}^{Z} & = & s (\beta \cos\theta)^2 (C_{ae}^2+C_{ve}^2) (C_{at}^2 + C_{vt}^2 ) + 8 s \beta \cos\theta C_{ae} C_{ve} C_{at} C_{vt} \nonumber\\&& + 
  (C_{ae}^2 + C_{ve}^2) \left((s - 4 m^2) C_{at}^2 + (s + 4 m^2) C_{vt}^2\right), \\
    f_{Z}(\kappa,\Lambda,\sqrt{s}) & = & (s - 4 m^2)\cos^{2}\theta (2\sin\theta_{w})^{2} \Biggl\{ (C_{at}^2 + C_{vt}^2 ) 2 \kappa C_{ve} \nonumber\\&& 
   -8 \kappa s \beta \cos\theta C_{at} C_{ve} (2\sin\theta_{w})^{2} \sin^{2}\left(\frac{p_{4}\Theta p_{3}}{2}\right)\nonumber
  \end{eqnarray}
  $$  +(C_{ae}^2 + C_{ve}^2)( (2 \kappa \sin\theta_{w})^{2} \sin^{2}\left(\frac{p_{4}\Theta p_{3}}{2}\right) + 2 \kappa C_{vt} ) \Biggr\}.$$
The photon and the $Z$ boson mediated diagrams interfere and contributes to the 
differential cross section given by     
 \begin{eqnarray}
  \frac{d\sigma^{\gamma Z}}{d\Omega} & = & \left(\frac{\alpha^2 \beta A_{f} }{16 s \sin^{2}2\theta_{w}((s-m_{Z})^{2}+(\Gamma_{Z} m_{z})^{2})} \right) \nonumber\\&&
\Biggl\{ 2 (s-m_{Z})^{2} Q_{e} Q_{t} \left( 2 s \beta \cos\theta  C_{ae} C_{at} +  \left((1+\cos^{2}\theta)+\left(\frac{4 m^2}{s}\right)\sin^{2}\theta \right) C_{ve} C_{vt} \right)\nonumber\\&& + 
f_{\gamma Z}(\kappa,\Lambda,\sqrt{s}) \Biggr\}
\end{eqnarray}
The term which contains the noncommutative correction is given by
 \begin{eqnarray}
  f_{\gamma Z}(\kappa,\Lambda,\sqrt{s}) & = &  4 Q_{e} \kappa s \beta \cos\theta C_{ae} C_{at} \sin\left(\frac{p_{4}\Theta p_{3}}{2}\right) \Biggl\{ 2 (s-m_{Z})^{2} \sin\left(\frac{p_{4}\Theta p_{3}}{2}\right) 
  \nonumber\\&& -2 (m_{Z} \Gamma_{Z}) \cos\left(\frac{p_{4}\Theta p_{3}}{2}\right) \Biggr\} \nonumber\\&& 
  + s \left((1+\cos^{2}\theta)+\left(\frac{4 m^2}{s}\right)\sin^{2}\theta \right) C_{ve} Q_{e} \sin\left(\frac{p_{4}\Theta p_{3}}{2}\right)  \nonumber\\&& 
  \Biggl\{ 2 (s-m_{Z})^{2} \sin\left(\frac{p_{4}\Theta p_{3}}{2}\right) \left[ 2 \kappa C_{vt} - \kappa (2\sin\theta_{w})^{2} (2 \kappa+Q_{t})\right]  \nonumber\\&&
  - 2 (m_{Z} \Gamma_{Z}) \cos\left(\frac{p_{4}\Theta p_{3}}{2}\right)\left[  \kappa Q_{t}(2\sin\theta_{w})^{2} - 2\kappa C_{vt} \right] \Biggr\} \label{diffphotonZ}
 \end{eqnarray} 
Here $m$ is the top quark mass and $A_{f}=C_{f}\times 4 \times 0.389\times 10^{9}$. 
The factor 
$C_{f}(=3)$ is the quark color factor and other one is a $\rm{GeV^{-2}}$ to $\rm{pb}$ conversion factor.
\subsection{NC matrix elements square in the helicity amplitude technique}\label{app:helincxsec}
Here we present the NC matrix elements square in the helicity amplitude analysis. We can calculate the helicity differential cross section by using the Feynman rule  \ref{photonfeyn} 
and \ref{zbosonfeyn}.  For the photon ($\gamma$) mediated $t~\overline{t}$ process, the differential cross section
\begin{eqnarray}
\frac{d \sigma^{\gamma}}{d \Omega} = \frac{\beta {\mathcal{A}}_{f}}{64 \pi^{2} s}\quad \sum_{h_{e^{-}},h_{e^{+}},\,h_{t},\,h_{\overline{t}}} |\mathcal{M}_{\gamma}|^{2}_{h_{e^{-}}h_{e^{+}}\, h_{t} \, h_{\overline{t}}}
\end{eqnarray}
Here ${\mathcal{A}}_{f} =C_{f}\times 1 \times 0.389\times 10^{9}$. $h_{f}~(=\lambda)$ is the helicity of the fermions $f= e^{\pm}$ and $t,\, \overline{t}$. 
The amplitude square $|{\mathcal{M}}_{\gamma}|^{2}_{h_{e^{-}}h_{e^{+}}\, h_{t} \, h_{\overline{t}}}$ for 
different helicity combination of the initial and final state particles are given by,
\begin{eqnarray}
 |\mathcal{M}_{\gamma}|^{2}_{LRLL} & = & |\mathcal{M}_{\gamma}|^{2}_{LRRR} = |\mathcal{M}_{\gamma}|^{2}_{RLLL} = |\mathcal{M}_{\gamma}|^{2}_{RLRR} = {\mathcal{F}}  \sin^{2}\theta \label{helistart} \\
  |\mathcal{M}_{\gamma}|^{2}_{LRLR} & = &  |\mathcal{M}_{\gamma}|^{2}_{RLLR}  = {\mathcal{F}}  \left( \frac{1+\cos\theta}{2} \right)^{2} \\
 |\mathcal{M}_{\gamma}|^{2}_{LRRL} & = & |\mathcal{M}_{\gamma}|^{2}_{RLRL} =  {\mathcal{F}}  \left( \frac{1-\cos\theta}{2} \right)^{2}
\end{eqnarray}
Here ${\mathcal{F}} = 64 \pi^{2} \alpha^{2} m^{2}\,Q_{e}^{2} \left( Q_{t}^{2}+f^{h}_{\gamma}(\kappa, \Lambda, \sqrt{s} )\right)$, $L$ and $R$ represents the left-handed and the right-handed helicity of the scattering particles 
(i.e. initial and final state particles). 
For the $Z$ boson mediated $t~\overline{t}$ process, the differential cross section 
$$ \frac{d \sigma^{Z}}{d \Omega} = \frac{\beta \mathcal{A}_{f}}{64 \pi^{2} s\, \sin^{4}2\theta_{w}[(s-m_{Z}^{2})^{2}+(m_{Z}\Gamma_{Z})^{2}]}\quad \sum_{h_{e^{-}},h_{e^{+}},\,h_{t},\,h_{\overline{t}}} |\mathcal{M}_{Z}|^{2}_{h_{e^{-}}h_{e^{+}}\, h_{t} \, h_{\overline{t}}}$$
The amplitude square $|\mathcal{M}_{Z}|^{2}_{h_{e^{-}}h_{e^{+}}\, h_{t} \, h_{\overline{t}}}$ 
for different helicity combination of the initial and final state particles are given by,
\begin{eqnarray}
 |\mathcal{M}_{Z}|^{2}_{LRLL} & = & |\mathcal{M}_{Z}|^{2}_{LRRR} = 64 \pi^{2} \alpha^{2} m^{2}\,(4s) (g_{lez})^{2}\left( C_{vt}^{2}+f^{h}_{Z\,0}(\kappa, \Lambda, \sqrt{s} )\right) \sin^{2}\theta \\
 |\mathcal{M}_{Z}|^{2}_{RLLL} & = & |\mathcal{M}_{Z}|^{2}_{RLRR} = 64 \pi^{2} \alpha^{2} m^{2}\,(4s) (g_{rez})^{2}\left( C_{vt}^{2}+f^{h}_{Z\,0}(\kappa, \Lambda, \sqrt{s} )\right) \sin^{2}\theta \\
 |\mathcal{M}_{Z}|^{2}_{LRLR} & = &  64 \pi^{2} \alpha^{2} \,(s) (g_{lez})^{2} (1+\cos\theta)^{2} \Biggr\{ C_{at}(C_{at}(s-4m^{2})+C_{vt} \, s \, \beta) \nonumber \\ && +s\, C_{vt} (C_{vt}+C_{at} \, \beta) + f^{h}_{Z\,+}(\kappa, \Lambda, \sqrt{s} )\Biggr\} \\
 |\mathcal{M}_{Z}|^{2}_{RLLR} & = &  64 \pi^{2} \alpha^{2} \,(s) (g_{rez})^{2} (1+\cos\theta)^{2} \Biggr\{ C_{at}(C_{at}(s-4m^{2})+C_{vt} \, s \, \beta) \nonumber \\ && +s\, C_{vt} (C_{vt}+C_{at} \, \beta) + f^{h}_{Z\,+}(\kappa, \Lambda, \sqrt{s} )\Biggr\} \\ 
 |\mathcal{M}_{Z}|^{2}_{LRRL} & = &  64 \pi^{2} \alpha^{2} \,(s) (g_{lez})^{2} (1-\cos\theta)^{2} \Biggr\{ C_{at}(C_{at}(s-4m^{2})-C_{vt} \, s \, \beta) \nonumber \\ && +s\, C_{vt} (C_{vt} - C_{at} \, \beta) + f^{h}_{Z\,-}(\kappa, \Lambda, \sqrt{s} )\Biggr\} \\
 |\mathcal{M}_{Z}|^{2}_{RLRL} & = &  64 \pi^{2} \alpha^{2} \,(s) (g_{rez})^{2} (1-\cos\theta)^{2} \Biggr\{ C_{at}(C_{at}(s-4m^{2})-C_{vt} \, s \, \beta) \nonumber \\ && +s\, C_{vt} (C_{vt} - C_{at} \, \beta) + f^{h}_{Z\,-}(\kappa, \Lambda, \sqrt{s} )\Biggr\}  
\end{eqnarray}
Finally, the interference term arising from the $\gamma$ and $Z$ boson mediated diagrams, is 
\begin{equation}
 \frac{d \sigma^{\gamma Z}}{d \Omega} = \frac{\beta \mathcal{A}_{f}}{64 \pi^{2} s\, \sin^{2}2\theta_{w}[(s-m_{Z}^{2})^{2}+(m_{Z}\Gamma_{Z})^{2}]}\quad \sum_{h_{e^{-}},h_{e^{+}},\,h_{t},\,h_{\overline{t}}} |\mathcal{M}_{\gamma Z}|^{2}_{h_{e^{-}}h_{e^{+}}\, h_{t} \, h_{\overline{t}}}
\end{equation}
The amplitude square $|\mathcal{M}_{\gamma Z}|^{2}_{h_{e^{-}}h_{e^{+}}\, h_{t} \, h_{\overline{t}}}$ 
for different helicity combination of the initial and final state particles are given by, 
\begin{eqnarray}
 |\mathcal{M}_{\gamma Z}|^{2}_{LRLL} & = & |\mathcal{M}_{\gamma Z}|^{2}_{LRRR} = 64 \pi^{2} \alpha^{2} 4m^{2} \, Q_{e}\, g_{lez}  \sin^{2} \theta \Biggr\{  (s-m^{2}_{Z})Q_{t}C_{vt} \nonumber \\ && + f^{h}_{\gamma Z\,0}(\kappa, \Lambda, \sqrt{s} )\Biggr\}
    \end{eqnarray}
   \begin{eqnarray}
 |\mathcal{M}_{\gamma Z}|^{2}_{RLLL} & = & |\mathcal{M}_{\gamma Z}|^{2}_{RLRR}  = 64 \pi^{2} \alpha^{2} 4m^{2} \, Q_{e} \, g_{rez} \sin^{2}\theta  \Biggr\{ (s-m^{2}_{Z})Q_{t}C_{vt} \nonumber \\ && +f^{h}_{\gamma Z\,0}(\kappa, \Lambda, \sqrt{s} )\Biggr\} \\
 |\mathcal{M}_{\gamma Z}|^{2}_{LRLR} & = &  64 \pi^{2} \alpha^{2} \,(s) Q_{e} \, g_{lez}\, (1+\cos\theta)^{2} \Biggr\{ (s-m^{2}_{Z})Q_{t}(C_{vt}+ C_{at} \, \beta) \nonumber \\ && - f^{h}_{\gamma Z\,+}(\kappa, \Lambda, \sqrt{s} )\Biggr\} \\
 |\mathcal{M}_{\gamma Z}|^{2}_{RLLR} & = &  64 \pi^{2} \alpha^{2} \,(s) Q_{e} \, g_{rez}\, (1+\cos\theta)^{2} \Biggr\{ (s-m^{2}_{Z})Q_{t}(C_{vt}+ C_{at} \, \beta) \nonumber \\ && - f^{h}_{\gamma Z\,+}(\kappa, \Lambda, \sqrt{s} )\Biggr\} \\
 |\mathcal{M}_{\gamma Z}|^{2}_{LRRL} & = &  64 \pi^{2} \alpha^{2} \,(s) Q_{e} \, g_{lez}\, (1-\cos\theta)^{2} \Biggr\{ (s-m^{2}_{Z})Q_{t}(C_{vt} - C_{at} \, \beta) \nonumber \\ && - f^{h}_{\gamma Z\,-}(\kappa, \Lambda, \sqrt{s} )\Biggr\} \\ 
 |\mathcal{M}_{\gamma Z}|^{2}_{RLRL} & = &  64 \pi^{2} \alpha^{2} \,(s) Q_{e} \, g_{rez}\, (1-\cos\theta)^{2} \Biggr\{ (s-m^{2}_{Z})Q_{t}(C_{vt} - C_{at} \, \beta) \nonumber \\ && - f^{h}_{\gamma Z\,-}(\kappa, \Lambda, \sqrt{s} )\Biggr\} 
\end{eqnarray}
We see that the noncommutative contribution for the top pair production at the linear collider 
as calculated by using the helicity amplitude techniques, solely depends on the following VSR sub-group 
$T(2)$ invariant functions $ f^{h}_{\gamma},f^{h}_{Z0},f^{h}_{Z\pm},f^{h}_{\gamma Z0}, f^{h}_{\gamma Z\pm}$ which are defined below: 
\begin{eqnarray}
 f^{h}_{\gamma}(\kappa, \Lambda, \sqrt{s} ) & =  & 2 (2\kappa)(\kappa+Q_{t}) \sin^{2}\left( \frac{p_{4}\Theta p_{3}}{2}\right) \\
 f^{h}_{Z0}(\kappa, \Lambda, \sqrt{s} ) & =  & (2\sin\theta_{w})^{2} \sin^{2}\left(\frac{p_{4}\Theta p_{3}}{2}\right) \left[ (2\kappa \sin^{2}\theta_{w})^{2}+2\kappa C_{vt}\right] \\
 f^{h}_{Z\pm}(\kappa, \Lambda, \sqrt{s} ) & =  & s (2\sin\theta_{w})^{2} \sin\left(\frac{p_{4}\Theta p_{3}}{2}\right) \Biggl\{ \kappa \, C_{vt} \nonumber \\ && + \sin\left(\frac{p_{4}\Theta p_{3}}{2}\right) [ (2\, \kappa \sin\theta_{w})^{2} \mp 2\, \kappa \, C_{at} ]\Biggr\} \\
  f^{h}_{\gamma Z0}(\kappa, \Lambda, \sqrt{s} ) & =  &(s-m^{2}_{Z}) \sin\left(\frac{p_{4}\Theta p_{3}}{2}\right) 
 \Biggl\{2 \sin\left(\frac{p_{4}\Theta p_{3}}{2}\right) (2 \kappa \sin\theta_{w})^{2}  
 \nonumber \\ && +  \kappa \, \cos\left(\frac{p_{4}\Theta p_{3}}{2}\right) [ 2\, C_{vt} + Q_{t}(2\, \sin\theta_{w})^{2} ]\Biggr\} \nonumber
  \end{eqnarray}
   \begin{eqnarray}
  && +  (m_{Z}\Gamma_{Z})\kappa \, \sin^{2}\left(\frac{p_{4}\Theta p_{3}}{2}\right) [ 2\, C_{vt} - Q_{t}(2\, \sin\theta_{w})^{2} ] \\
 f^{h}_{\gamma Z\pm}(\kappa, \Lambda, \sqrt{s} ) & =  &   (s-m^{2}_{Z}) \sin^{2}\left(\frac{p_{4}\Theta p_{3}}{2}\right) 
 \Biggl\{2 (2 \kappa \sin\theta_{w})^{2} + Q_{t}\, \kappa \, (2\sin\theta_{w})^{2} 
 \nonumber \\ && +  2\, \kappa \,(  C_{vt} \pm C_{at} \, \beta )\Biggr\} +  (m_{Z}\Gamma_{Z}) \sin\left(\frac{p_{4}\Theta p_{3}}{2}\right) \cos\left(\frac{p_{4}\Theta p_{3}}{2}\right) \nonumber \\ &&  
 \Biggl[  2\kappa\,(C_{vt} \pm C_{at}\, \beta) + \kappa\, Q_{t}(2\, \sin\theta_{w})^{2} \Biggr] \label{heliend}
\end{eqnarray}
It is worthwhile to note that the set of equations from \ref{helistart} to  \ref{heliend} are, in principle,
applicable to any generic fermion pair production $ e^{-}e^{+}  \xrightarrow{\gamma/Z} f \, \overline{f}$ at the Linear Collider. 
In case the final particles are leptons, one need to replace $ C_{vt},C_{at},Q_{t}$ and $m$ by $ C_{vl},C_{al},
Q_{l},m_{l}$ and $C_{f} = 1$, respectively. Similarly, for quarks, one can substitute charge,mass and 
couplings of the quarks accordingly. Furthermore, these equations can easily be extended to
partonic level pair production at the LHC as well. For gluon calculation, one has to compute proper polarization contribution.
Assuming the partons to be 
massless, we can substitute $Q_{e}=Q_{quark},\, g_{lez}=g_{lqz}$ and $g_{rez}=g_{rqz}$ in above 
equations (\ref{helistart}-\ref{heliend}). Here we take $$ g_{lfz}= T^{f}_{3}-Q_{f} \sin^{2}\theta_{w} \quad \quad g_{rfz}= -Q_{f}\sin^{2}\theta_{w} $$
Finally, one needs to substitute the proper average color factor as well. 

\acknowledgments
The work of SJ and PK is supported by Physical Research Laboratory (PRL), Department of Space, Government of India and also acknowledge
the computational  support from Vikram-100 HPC at PRL. Authors would like to thank Prof. Josip Trampetic, The Ruder Boskovic Institute, Zagreb, Croatia for useful discussions and 
various comments on UV/IR mixing. Further SJ would like to thank Prof.V. Ravindran, Institute of Mathematical Sciences, Chennai, India for fruitful discussions about HQET parametrization and UV/IR mixing.
PKD thanks the Department of Physics and Astronomy, University of Kansas, USA for providing the local hospitality 
where a part of the work was completed. The work of PKD is partly supported by the SERB Grant No. EMR/2016/002651.



\end{document}